\documentclass{article} 
\usepackage{graphicx}	
\usepackage{amsmath}	
\usepackage{amssymb}	
\usepackage{authblk}

\newcommand{\beq}{\begin{equation}}
\newcommand{\eeq}{\end{equation}}
\newcommand{\beqa}{\begin{eqnarray}}
\newcommand{\eeqa}{\end{eqnarray}}

\title{Numerical Simulations 
of Two-Fluid Magnetoacoustic Waves in the Solar Atmosphere}

\author[1]{J. Kra\'skiewicz}
\author[1]{K. Murawski} 
\author[2]{Z.E.~Musielak}

\affil[1]{
Institute of Physics, University of Maria Curie-Sk{\l}odowska, Pl.\ Marii Curie-Sk{\l}odowskiej 5, 20-031 Lublin, Poland}
\affil[2]{
Department of Physics, University of Texas at Arlington, Arlington, TX 76019, USA
}

\begin{document}
\maketitle

\begin{abstract}
We study vertical variations of wave-periods of magnetoacoustic two-fluid 
waves in the partially ionized lower solar atmosphere, consisting of ion (proton) + electron and neutral (atomic hydrogen) fluids,
which are coupled by ion-neutral collisions.  The study allows finding the wave period cutoffs and their 
variations in the solar atmosphere, as well as establishing the role of these cutoffs in determining 
the wave propagation conditions.  The atmosphere is permitted by a uniform vertical magnetic field.
We perform numerical simulations in the framework of a one-dimensional (1D), two-fluid model in 
which plane waves are exited by a harmonic driver in the vertical ion and neutral velocities, operating 
at the bottom of the solar photosphere.
We observe excitation of waves with cutoff wave-periods in addition to waves set directly by the driver. We also see that some 
waves exited by that driver can reach the solar corona.
Despite of its limitations such as the lack of non-adiabatic 
and non-ideal terms and a simple 1D structure, 
the developed two-fluid model of the solar atmosphere 
sheds a new light on the role of cutoffs 
in setting up the wave propagation conditions in the solar atmosphere and finding periods of waves that may carry their energy from the solar surface to the corona. 
\end{abstract}

Methods: numerical -- Sun: chromosphere -- Sun: transition region


%
%
\section{INTRODUCTION}
%
%
The structure of the solar atmosphere varies from partially ionized in its lower layers to fully 
ionized in the solar corona.  In the solar photosphere there is one ion per about $10^3-10^4$ 
neutrals but in the chromosphere and in the transition region the number of neutrals rapidly 
falls off with height as the temperature increases (e.g., Priest 2014; Ballester et al. 2018).  The 
role of wave heating in this temperature increase has been investigated in many papers (e.g., 
Narain \& Ulmschneider 1996; Roberts \& Ulmschneider 1997; Roberts 1991, 2006) in which 
a fully ionized solar atmosphere is typically considered. Recently, the effects of partially ionized solar atmosphere on the wave propagation 
are taken into account in two-fluid numerical studies performed by Maneva et al. (2017), W\'ojcik et al. 
(2019), Popescu Braileanu et al. (2019), Ku\'zma et al. (2019) and Murawski et al. (2020).  
The obtained results demonstrate that ion and neutral waves behave differently and that 
ion-neutral collisions may become an effective damping mechanism for the waves.

The wave propagation in the solar atmosphere is strongly affected by the presence of 
cutoff periods, which are different for different waves and strongly depend on the 
structure of media in which the waves propagate.  The concept of cutoff periods was 
originally introduced by Lamb (1909, 1910, 1945) for acoustic waves propagating in
stratified but isothermal atmospheres.   Then, the work was extended to more realistic 
(nonisothermal) atmospheres with magnetic fields in which the cutoffs for different 
waves were obtained analytically (e.g., Defouw 1976; Gough 1977; Rae \& Roberts
1982; Musielak 1990; Fleck \& Schmitz 1991, 1993; Schmitz \& Fleck 1992; Stark
\& Musielak 1993; Musielak \& Moore 1995; Roberts \& Ulmschneider 1997; Roberts 
2004, 2006; Musielak et al. 2006, 2007;  Routh et al. 2010; Murawski \& Musielak
2010; Cally \& Hasan 2011; Routh \& Musielak 2014; Perera et al. 2015; Felipe et
al. 2018; Routh et al. 2020).   All these analytical studies were done for fully ionized (and often isothermal) solar atmosphere; despite these limitations some of the derived cutoffs for acoustic and magnetoacoustic waves (MAWs) will be compared to the results for two-fluid MAWs obtained in this paper. 

The above analytical studies of cutoff periods were supplemented by 3D numerical
simulations of different linear and nonlinear waves in the solar atmosphere with
different non-magnetic and magnetic settings, and the wave generation, propagation 
and dissipation were investigated.  Specific numerical results involve impulsively 
generated linear and non-linear magnetohydrodynamic (MHD) waves, 3D simulations of magnetic twisters, 
and 3D simulations of magnetic flux tube waves (e.g., Murawski \& Zaqarashvili
2010; Chmielewski et al. 2013; Murawski et al. 2016; Mart{\'{\i}}nez-Sykora et al. 
2017; Maneva et al. 2017; Kraśkiewicz et al. 2019; Ku\'zma et al.\ 2019; 
Popescu Braileanu et al. 2019; W\'ojcik et al. 2018, 2019, 2020; Murawski et al.
2020).  Some of this work has already included the effects of partial ionization
of the solar atmosphere. 
Specifically, the observationally established variations 
of the acoustic cutoff periods with height in the solar atmosphere (Wi\'sniewska et al. 
2016; Kayshap et al. 2018) were partially reproduced numerically by Murawski 
\& Musielak (2016), Murawski et al. (2016); however, recently Ku\'zma et al. (2022)
obtained a very good agreement with the observations. 

The main purpose of this paper is to extend the MHD models (e.g., Fleck \& Schmitz 1991, Kalkofen et al. 1994, Kra\'skiewicz et al. 2019) 
according to which for the piston-wave excitation 
in the isothermal atmosphere, 
the asymptotic solution shows in the linear limit a superposition of waves with the piston wave-period and waves with the cutoff wave-period. 
For the waves excited above the cutoff period, the atmosphere oscillates with the acoustic cutoff period at greater heights; however, as time elapses the piston driving period takes over. 
The extension is done to account for a non-isothermal solar
atmosphere, the conditions for propagation of 
long wave-period two-fluid MAWs are determined by taking into account 
the effects of dynamics of neutrals, and by considering 
wave-periods of neutral waves.

The paper is organized as follows. In Sec.~\ref{sec:numerical_model} we describe 
the numerical model of the atmosphere. In Sec.~\ref{sec:num_sim_model} we 
present our numerical results. We finalize our draft by conclusions in Sec.~4. 
%
\section{NUMERICAL MODEL OF THE SOLAR ATMOSPHERE}\label{sec:numerical_model}
%
%
We consider a 1D magnetically structured 
and gravitationally stratified solar atmosphere 
which dynamics is described 
by set of two-fluid equations for ions + electrons treated as one fluid and neutrals regarded as second fluid. 
Specifically, these fluids are governed by the following equations (e.g., Maneva et al. 2017; Popescu Braileanu
et al. 2019): 
\begin{equation}
\frac{\partial \varrho_{\rm n}}{\partial t}+\nabla\cdot(\varrho_{\rm n} \mathbf{V}_{\rm n}) = 0\,,
\label{eq:neutral_continuity}
\end{equation}
\begin{equation}
\frac{\partial \varrho_{\rm i}}{\partial t}+\nabla\cdot(\varrho_{\rm i} \mathbf{V}_{\rm i}) = 0\,,
\label{eq:ion_continuity}
\end{equation}
\begin{equation}
\begin{split}
\frac{\partial (\varrho_{\rm n} \mathbf{V}_{\rm n})}{\partial t}+
\nabla \cdot (\varrho_{\rm n} \mathbf{V}_{\rm n} \mathbf{V}_{\rm n}+p_{\rm n} \mathbf{I}) = 
\varrho_{\rm n} \mathbf{g} + {\bf S_{\rm m}}\,,
\end{split}
\label{eq:neutral_momentum}
\end{equation}
\begin{equation}
\begin{split}
\frac{\partial  (\varrho_{\rm i} \mathbf{V}_{\rm i})}{\partial t}+
\nabla \cdot (\varrho_{\rm i} \mathbf{V}_{\rm i} \mathbf{V}_{\rm i}+p_{\rm i} \mathbf{I}) = 
\frac{1}{\mu}(\nabla \times \mathbf{B}) \times \mathbf{B}\\
+ \varrho_{\rm i} \mathbf{g} - {\bf S_{\rm m}}\,,
\end{split}
\label{eq:ion_momentum}
\end{equation}
\begin{equation}
\frac{\partial \mathbf{B}}{\partial t} = \nabla \times (\mathbf{V_{\rm i} \times }\mathbf{B})\,, \hspace{3mm} \nabla \cdot{\mathbf B}=0\,,
\label{eq:ions_induction}\\
\end{equation}
\begin{equation}
\begin{split}
\frac{\partial E_{\rm n}}{\partial t}+\nabla\cdot[(E_{\rm n}+p_{\rm n})\mathbf{V}_{\rm n}] =
 \varrho_{\rm n} \mathbf{g} \cdot \mathbf{V}_{\rm n} + S_{\rm En}\,,
\end{split}
\label{eq:neutral_energy}
\end{equation}
\begin{equation}
\begin{split}
\frac{\partial E_{\rm i}}{\partial t}+\nabla\cdot\left[\left(E_{\rm i}+p_{\rm i} + 
\frac{{\bf B}^2}{2\mu} \right)\mathbf{V}_{\rm i}-\frac{\mathbf{B}}{\mu}
(\mathbf{V}_{\rm i}\cdot \mathbf{B})\right]\\ = 
\varrho_{\rm i} \mathbf{g}  \cdot \mathbf{V}_{\rm i} + S_{\rm Ei}\,,
\end{split}
\label{eq:ion_energy}
\end{equation}
\begin{equation}
\begin{split}
E_{\rm n}=\frac{\varrho_{\rm n} \mathbf{V}_{\rm n}^2}{2}+\frac{p_{\rm n}}{\gamma-1}\,,
\end{split}
\label{eq:En}
\end{equation}
\begin{equation}
\begin{split}
E_{\rm i}=\frac{\varrho_{\rm n} \mathbf{V}_{\rm i}^2}{2}+\frac{{\bf B}^2}{2\mu}+\frac{p_{\rm i}}{\gamma-1}\,.
\end{split}
\label{eq:Ei}
\end{equation}
Here the momentum collisional, ${\bf S_{\rm m}}$, and energy, $S_{\rm Ei,n}$,
source terms are given as  
\begin{equation}
{\bf S_{\rm m}} =\alpha_{\rm c}({\bf V_{\rm i}}-{\bf V_{\rm n}})\, , 
\end{equation}
\begin{equation}
S_{\rm En}= 
\frac{1}{2}\alpha_{\rm c}(V_{\rm i}-V_{\rm n})^2
+
\frac{3\alpha_{\rm c} k_{\rm B}}{m_{\rm i}+m_{\rm n}}(T_{\rm i}-T_{\rm n})\, 
\end{equation}
%
\begin{equation}
S_{\rm Ei}= 
\frac{1}{2}\alpha_{\rm c}(V_{\rm i}-V_{\rm n})^2
+
\frac{3\alpha_{\rm c}k_{\rm B}}{m_{\rm i}+m_{\rm n}}(T_{\rm n}-T_{\rm i})\, 
\end{equation}
with subscripts $_{\rm i}$, $_{\rm n}$ and $_{\rm e}$ corresponding respectively to ions, neutrals and electrons. 
The symbols $\varrho_{\rm i,n}$ denote mass densities, ${\bf V}_{\rm i,n}=[V_{\rm i,n\,x}, V_{\rm i,n\,y}, 0]$ velocities, 
$p_{\rm i,n}$ ion+electron and neutral gas pressures, 
${\bf B}$ is magnetic field 
and $T_{\rm i,n}$ are temperatures specified by ideal gas laws, 
\begin{equation}
p_{\rm n} =\frac{k_{\rm B}}{m_{\rm n}}\varrho_{\rm n}T_{\rm n}\, , \hspace*{3mm} 
p_{\rm i}=\frac{k_{\rm B}}{m_{\rm i}}\varrho_{\rm i}T_{\rm i}\, .
\label{eq:pressures}
\end{equation}
%
\begin{figure}
\begin{center}
\includegraphics[width=8cm]{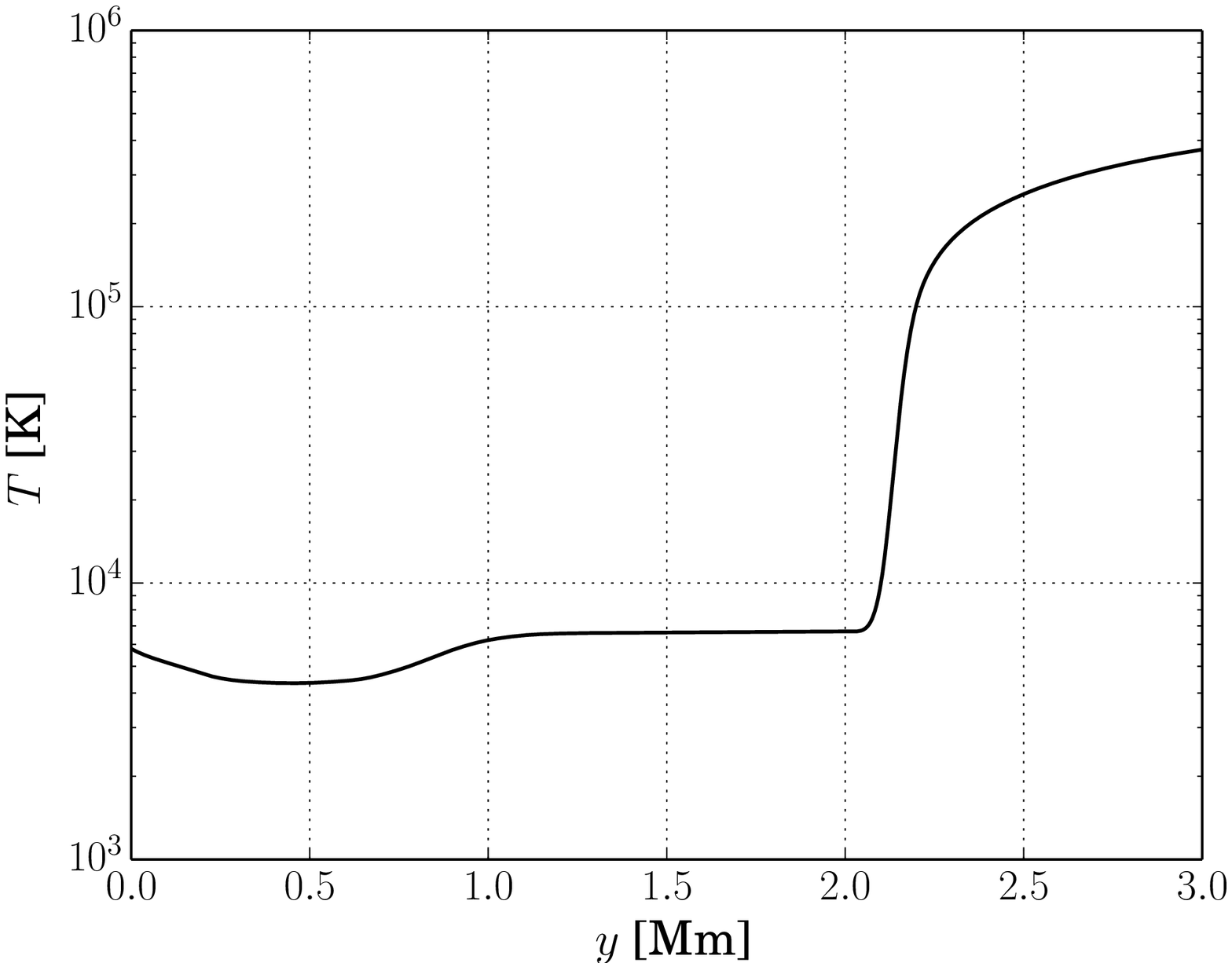}
\includegraphics[width=8cm]{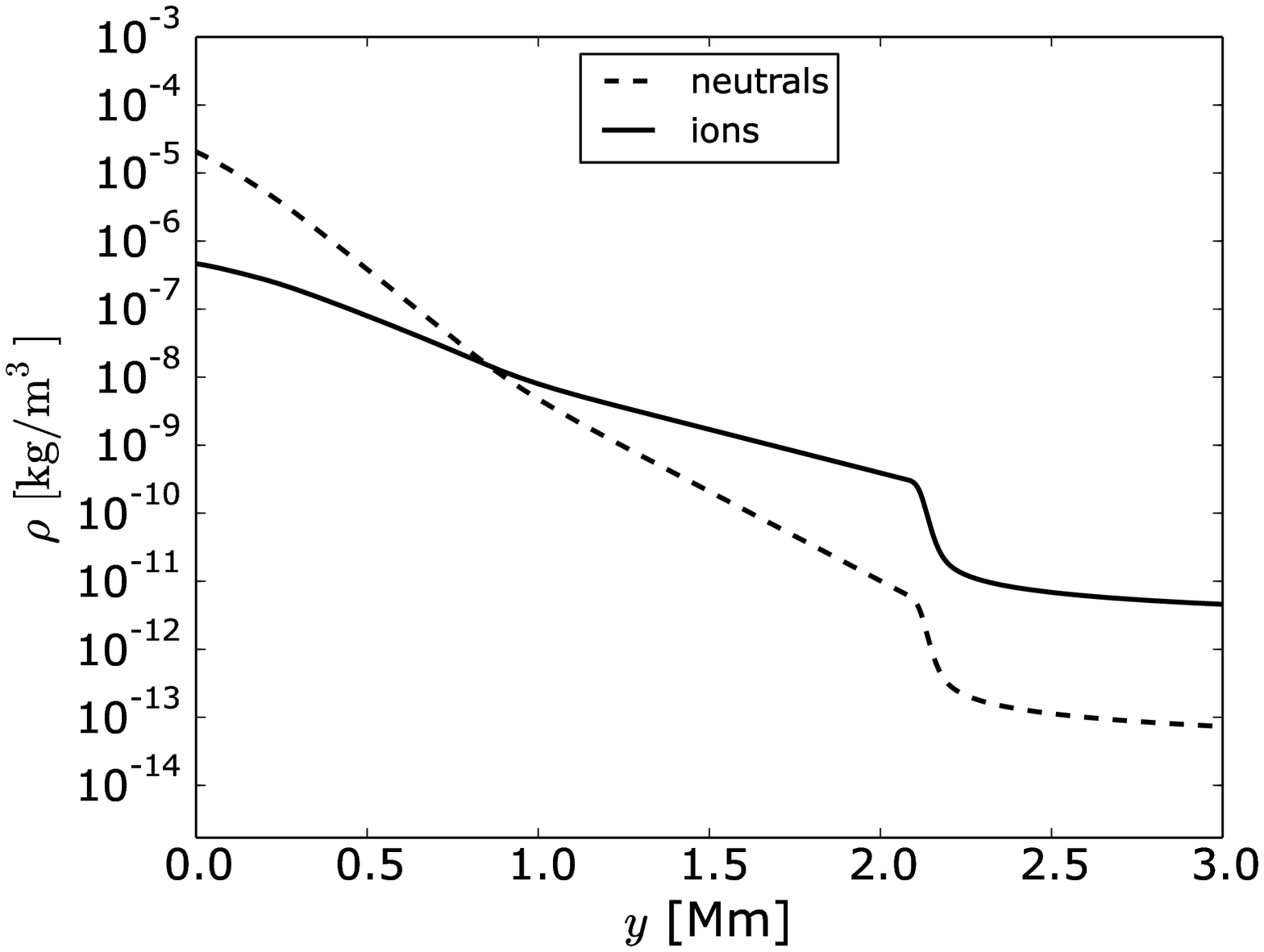}
\caption{\small 
Vertical profiles of equilibrium temperature (top) and 
the hydrostatic mass densities of 
the ionized (bottom, solid line) and neutral fluids (bottom, dashed line) 
versus height, $y$, in the two-fluid model of the solar atmosphere. 
}
\vspace{-1cm}
\label{fig:Atm}
\end{center}
\end{figure}  
Collision coefficient is given as (e.g. Oliver et al. 2016; Ballester et al. 2018, and references cited therein)
\begin{equation}
  \alpha_{\rm c} = 
  \frac{4}{3} 
 \frac{\sigma_{\rm in}}{m_{\rm i}+m_{\rm n}}
 \sqrt{ \frac{8k_{\rm B}}{\pi} 
\left(
\frac{T_{\rm i}}{m_{\rm i}}+\frac{T_{\rm n}}{m_{\rm n}}
\right) } \; \varrho_{\rm n} \varrho_{\rm i}
\end{equation}
with $\sigma_{\rm in}$ being the collisional cross-section 
for which we have chosen its quantum value of $1.4\times 10^{-19}$~m$^{2}$
(Vranjes and Krstic 2013).
\begin{figure} 
\begin{center}
\includegraphics[width=8cm]{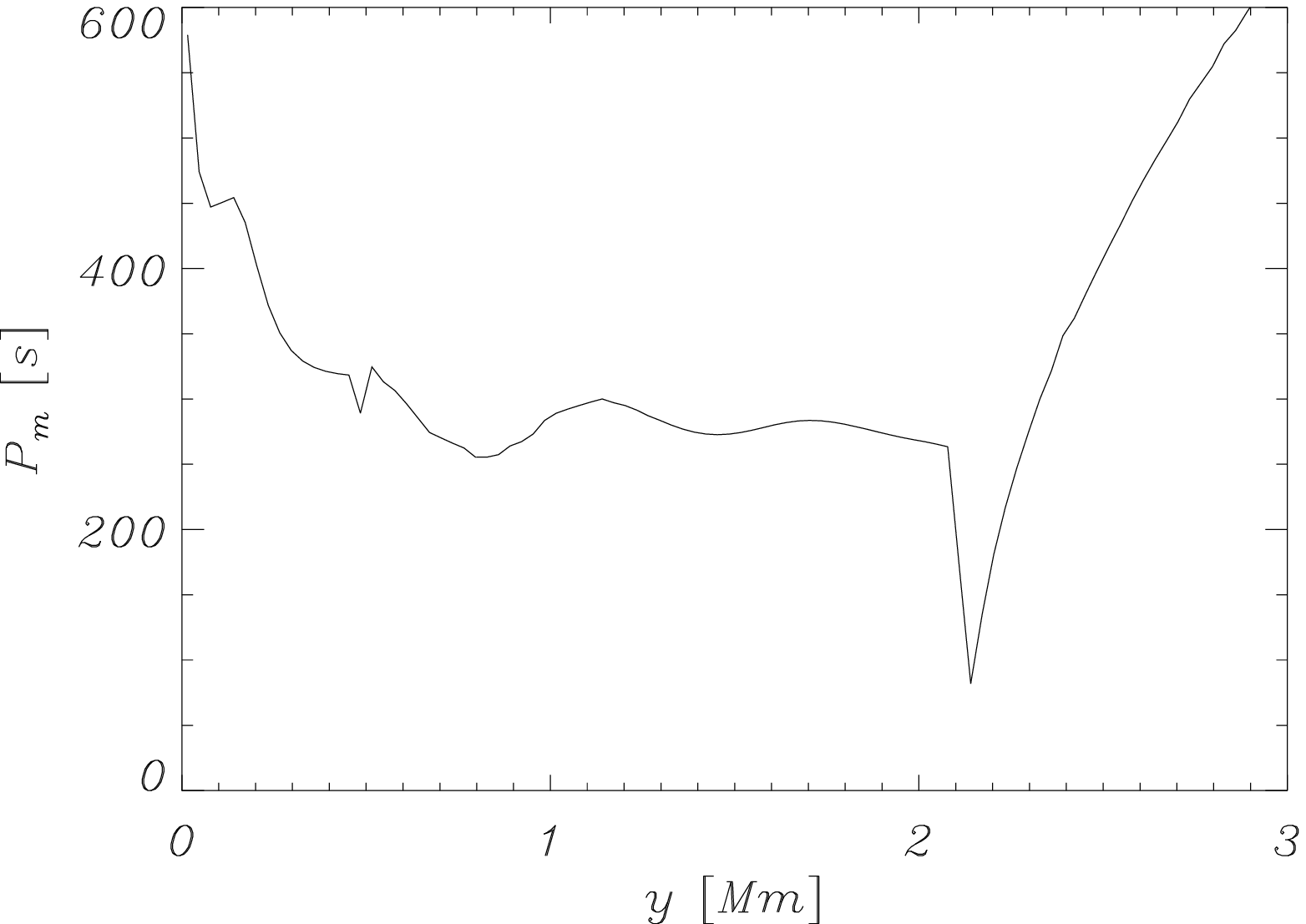}
\caption{\small The cutoff wave-period, $P_{\rm m}$, vs. height in the case of 
vertical magnetic field of magnitude $B_y=11.4$~G.
}
\label{fig:Pm_cut_off}
\end{center}
\end{figure}  
A gravity vector is ${\bf g} = [0, -g, 0]$ with its magnitude $g = 274.78$ m s$^{-2}$, 
%
$m_{\rm i,n}$
are the 
masses of respectively ions and neutrals, 
$k_{\rm B}$ is the Boltzmann constant, 
$\gamma=5/3$ is the specific heats ratio, and $\mu$ is magnetic permeability of the medium. 
The other symbols have their standard meaning. 

The present 1D model suffers from severe limitations. For instance,  important dissipative effects such as ionization/recombination, thermal conduction and radiative losses are neglected. This leads very likely to an overestimated energy flux of these waves that can be carried into the corona. In addition, the very significant magnetic expansion of flux tubes from the photosphere to the corona is neglected, leading to severe overestimation of the wave energy flux and the corresponding density perturbations. These effects are very important since the magnetic field strength could decrease by $2-3$ orders of magnitude  from the photosphere to the corona. 
We plan to make our model more realistic in future studies. 
%
%
\begin{figure} 
\begin{center}
\includegraphics[width=8cm]{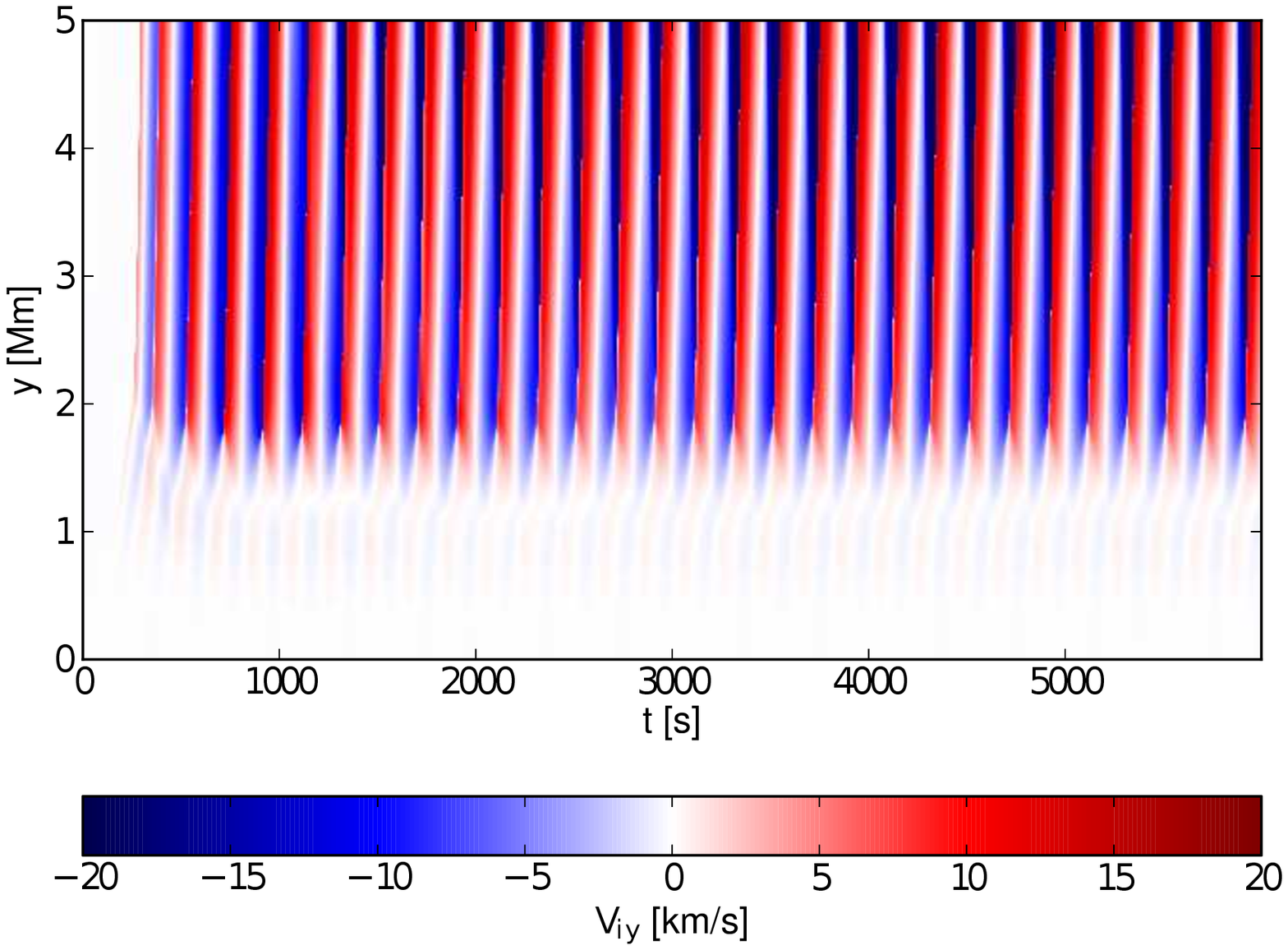}
\includegraphics[width=8cm]{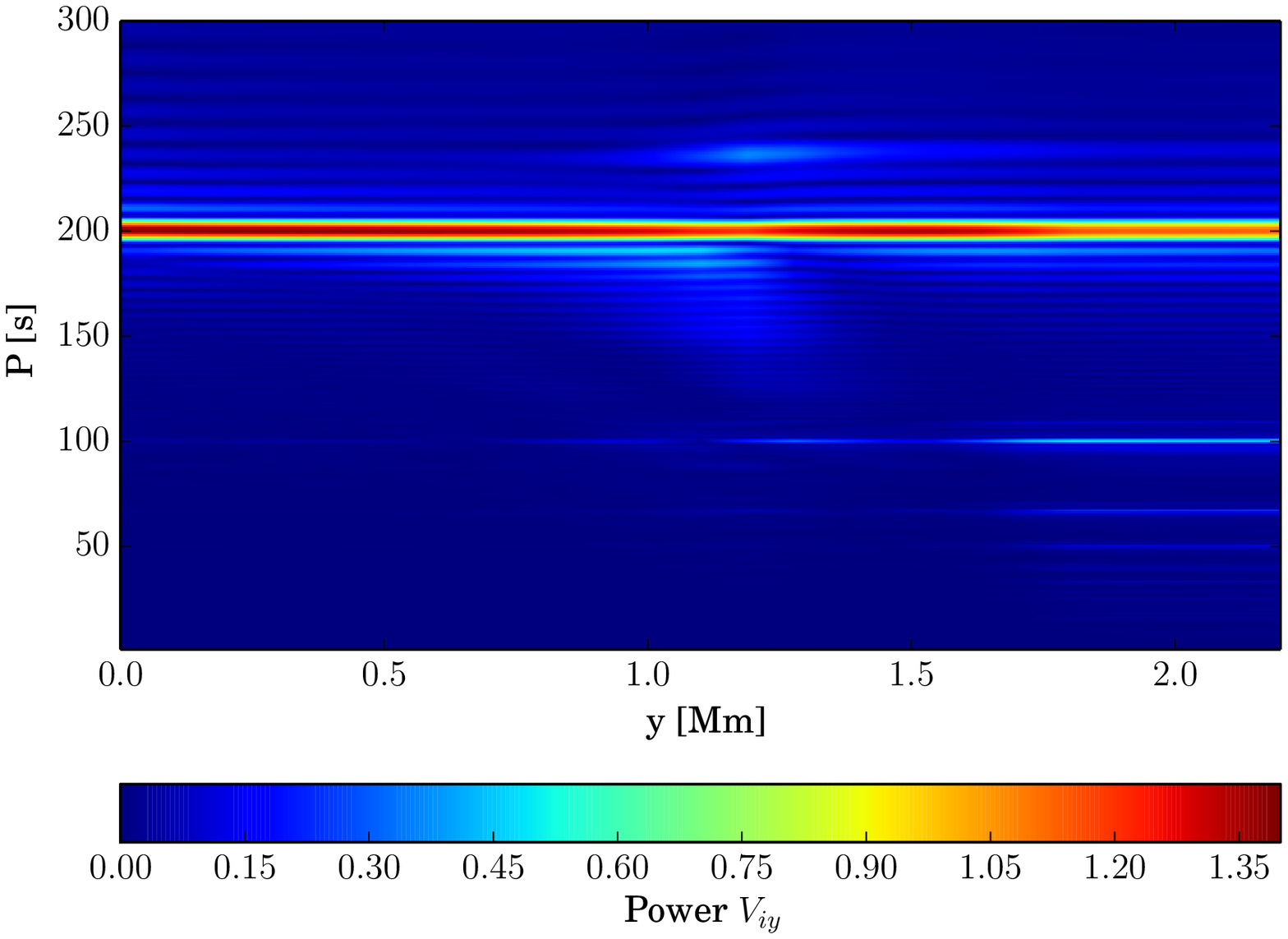}
\caption{\small Time-distance plot for the ion vertical velocity, $V_{\rm iy}$, (top) and its Fourier 
period, $P$, vs. height, $y$, (bottom) in the case of $P_{\rm d}=200$~s. 
}
\label{fig:200_Viy}
\end{center}
\end{figure}  
%
\begin{figure*} 
\begin{center}
\includegraphics[width=14cm, height=13cm]{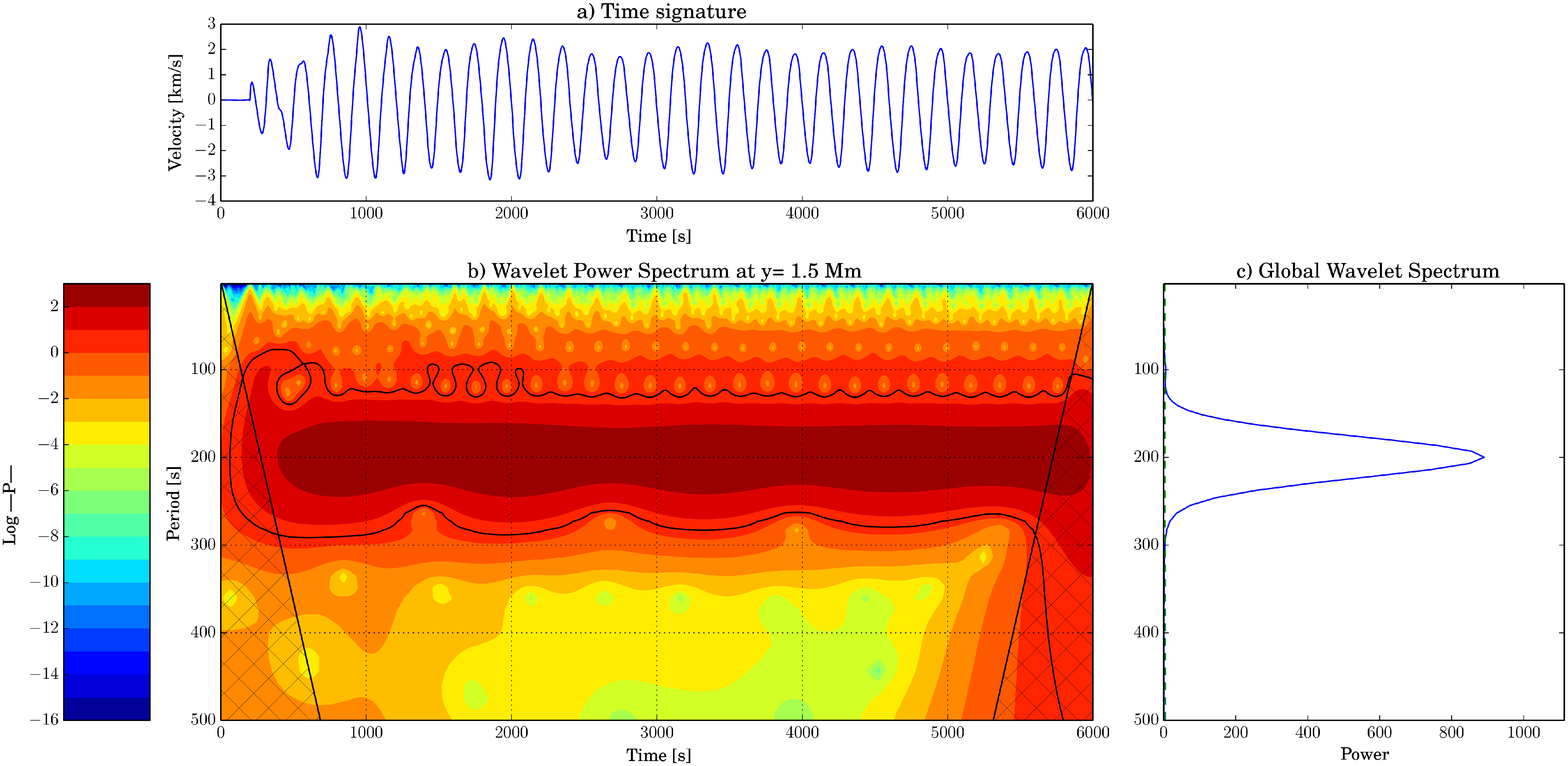}
\caption{\small Time-signature for $V_{\rm iy}(y=1.5\,{\rm Mm})$ (top), its wavelet spectrum (left-bottom), and global wavelet spectrum (right-bottom) in the case of $P_{\rm d}=200$~s. 
}
\label{fig:200_wavelet_Viy-1.5}
\end{center}
\end{figure*}  
\begin{figure} 
\begin{center}
\includegraphics[width=8cm]{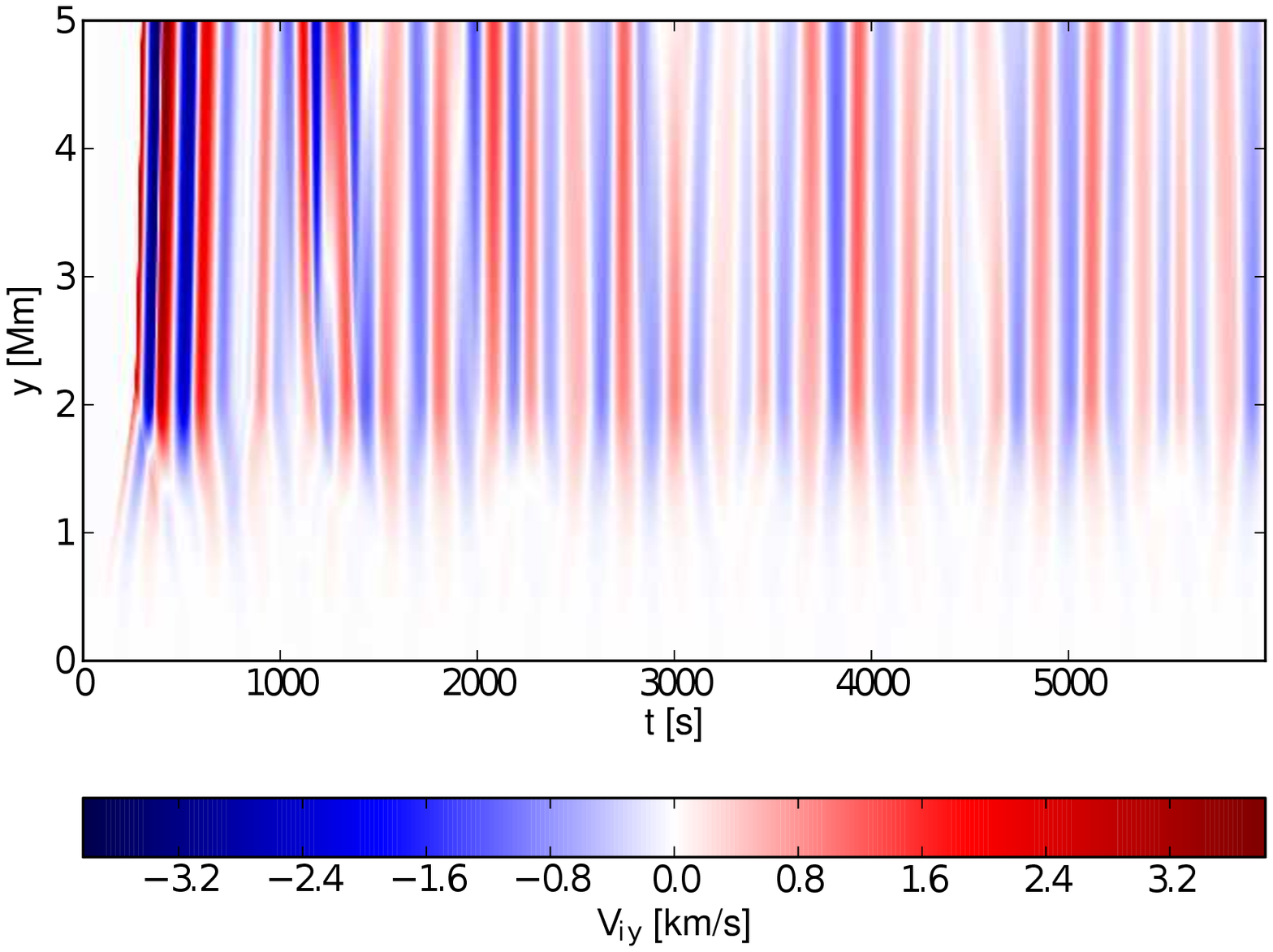}
\includegraphics[width=8cm]{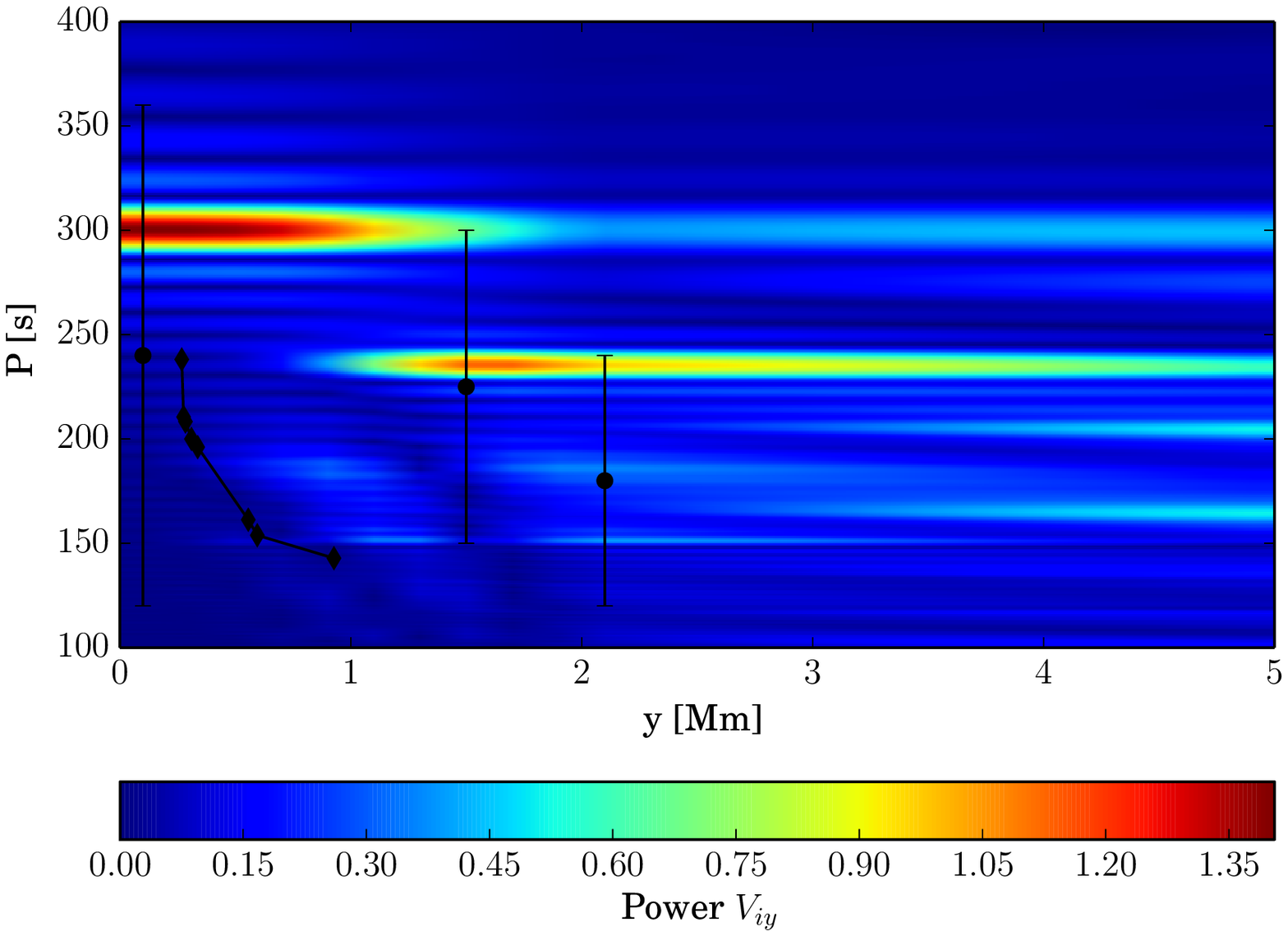}
\caption{\small Time-distance plot for the ion vertical velocity, $V_{\rm iy}$, (top) and its Fourier period, $P$, and experimental data from Wi\'sniewska et al. (2016) (stars) and Kayshap et al. (2018) (circles) vs. height, $y$, (bottom) in the case of $P_{\rm d}=300$~s.
}
\label{fig:300_Viy}
\end{center}
\end{figure}  
%
\begin{figure*} 
\begin{center}
\includegraphics[width=14cm, height=13cm]{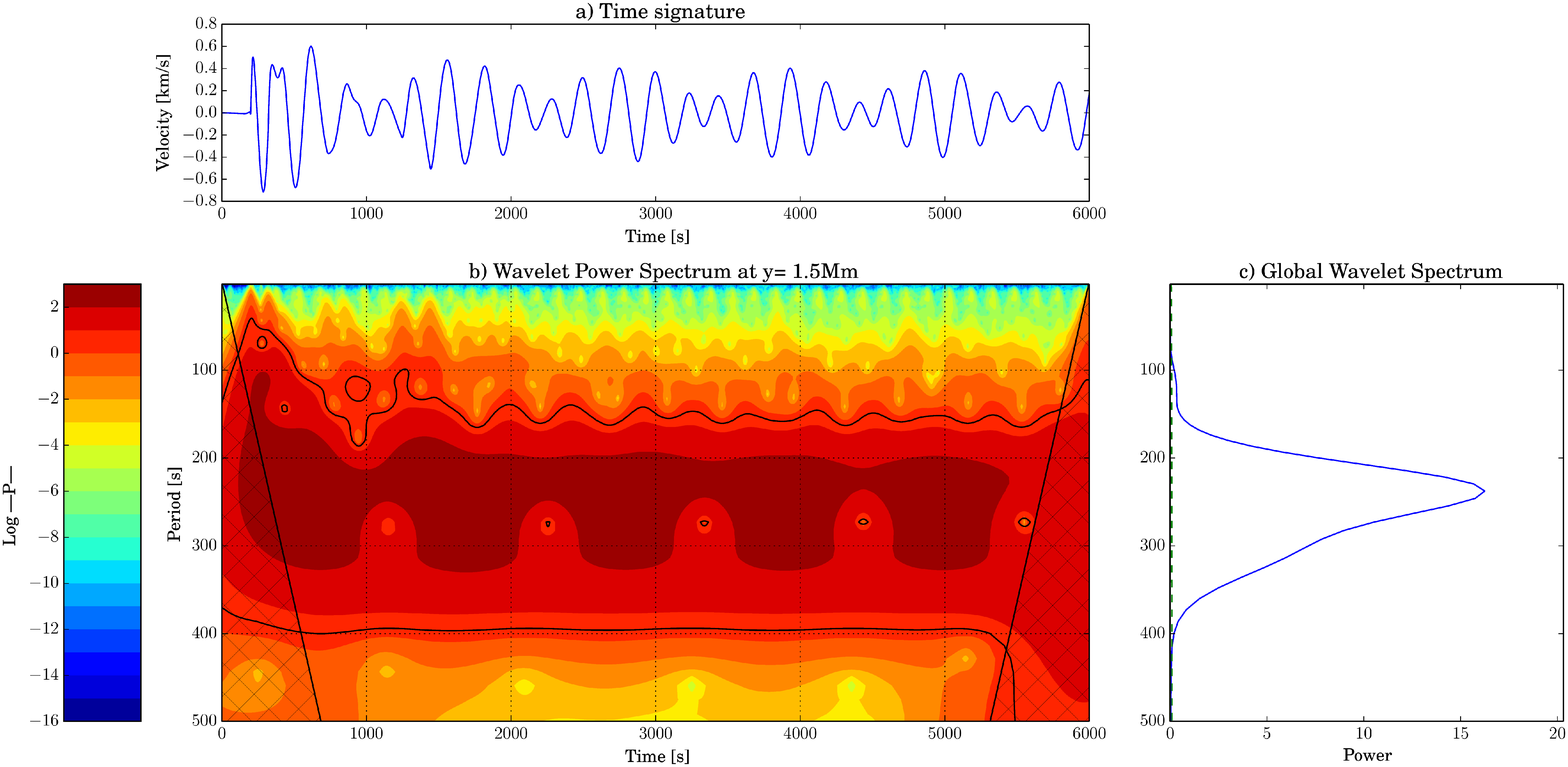}
\caption{\small Time-signature for $V_{\rm iy}(y=1.5\,{\rm Mm})$ (top), its wavelet 
spectrum (left-bottom), and global wavelet spectrum (right-bottom)  in the case of 
$P_{\rm d}=300$~s.
}
\label{fig:300_wavelet_Viy-1.5}
\end{center}
\end{figure*}  

\begin{figure*}
\begin{center}
\mbox{
\includegraphics[width=8cm]{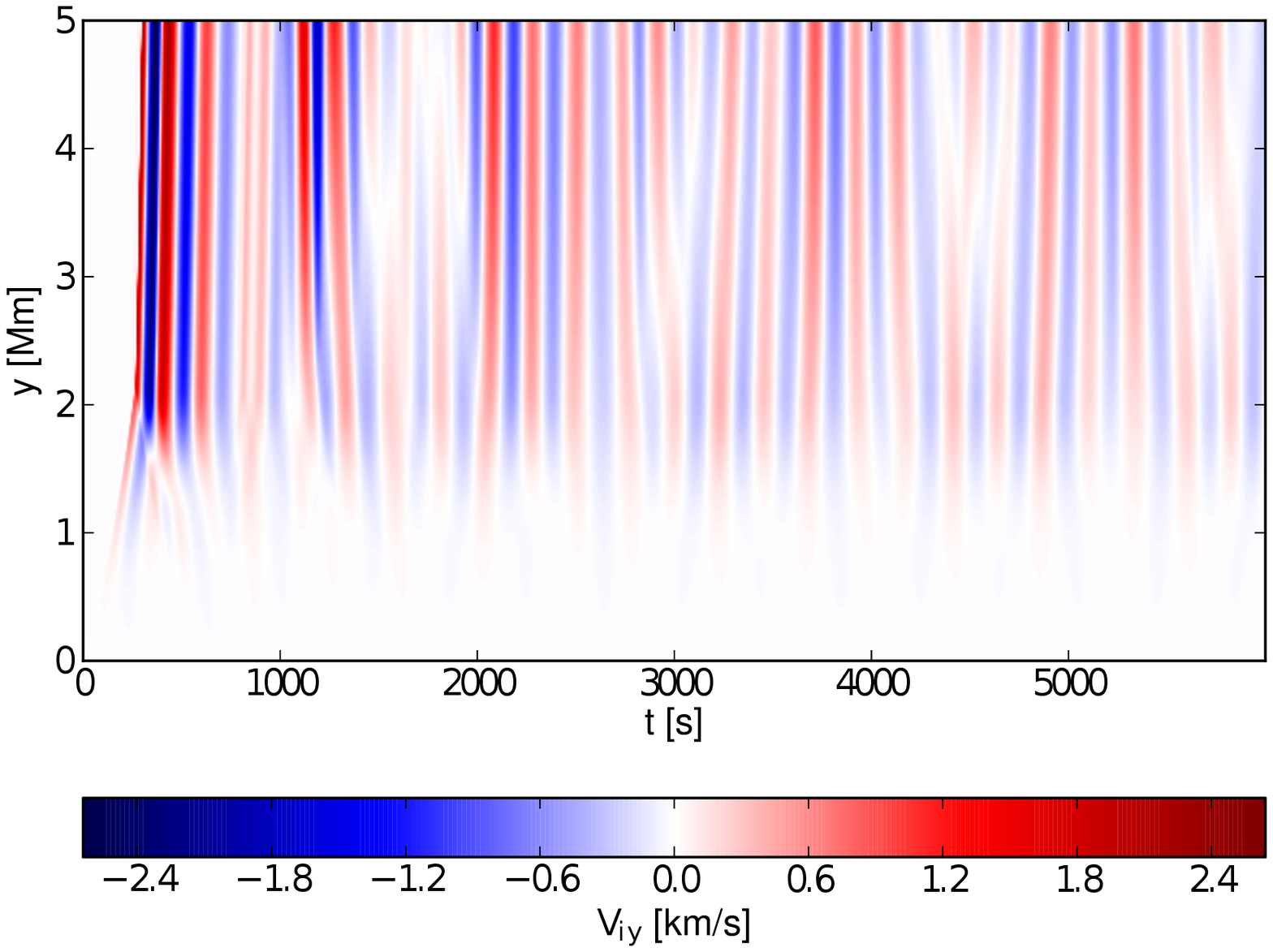}
\includegraphics[width=8cm]{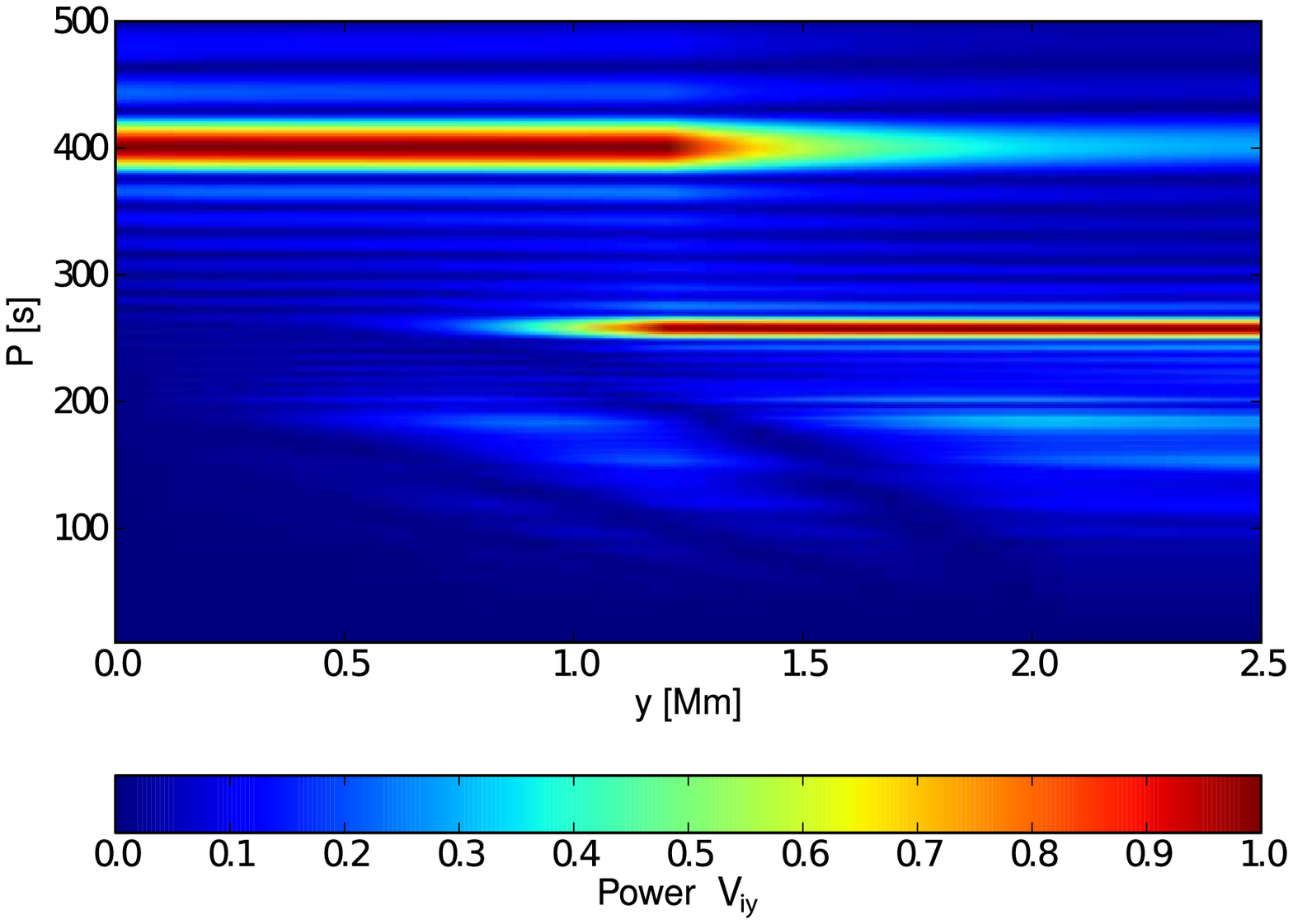}
}
\caption{\small Time-distance plots for the ion vertical velocity, $V_{iy}$, and its Fourier power period, $P$, vs. height, $y$, for $P_{\rm d}=400$~s.
}
\label{fig:400_Viy}
\end{center}
\end{figure*}  
\begin{figure*} 
\begin{center}
\includegraphics[width=14cm, height=11cm]{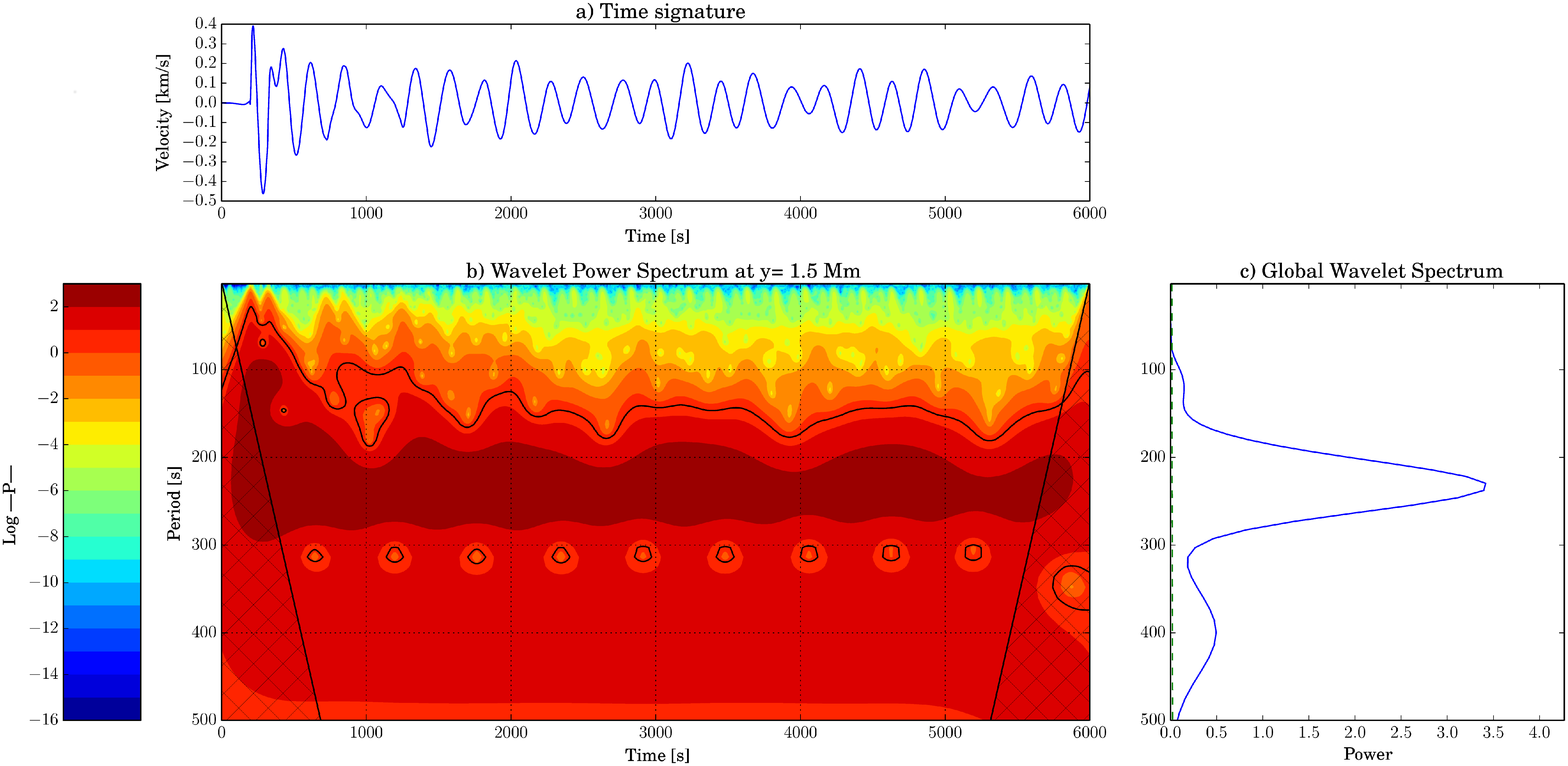}
\caption{\small Time-signature for $V_{\rm iy}(y=1.5\,{\rm Mm})$ (top), its wavelet spectrum (left-bottom), and global wavelet spectrum (right-bottom) in the case of $P_{\rm d}=400$~s.
}
\label{fig:wavelet_Pd400-1.5}
\end{center}
\end{figure*}  
\begin{figure} 
\begin{center}
\includegraphics[width=8cm]{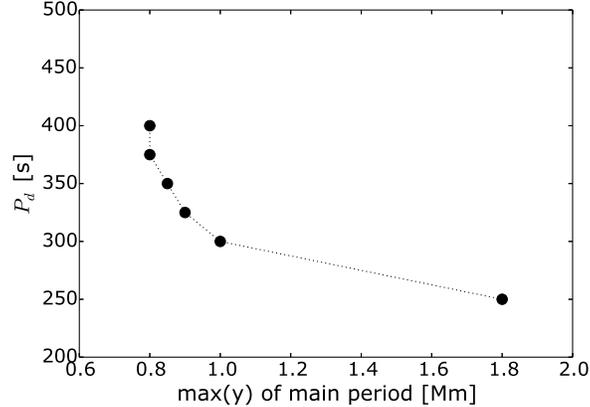}
\caption{\small 
Main Fourier period, $P_{\rm m}$, for $V_{\rm iy}$
vs. maximum value of $y$ at which this period is present 
in the case of $B_{\rm y}=11.4$~G.
}
\vspace{-0.5cm}
\label{fig:cutoff_Pd}
\end{center}
\end{figure}  
\begin{figure} 
\begin{center}
\includegraphics[width=8cm]{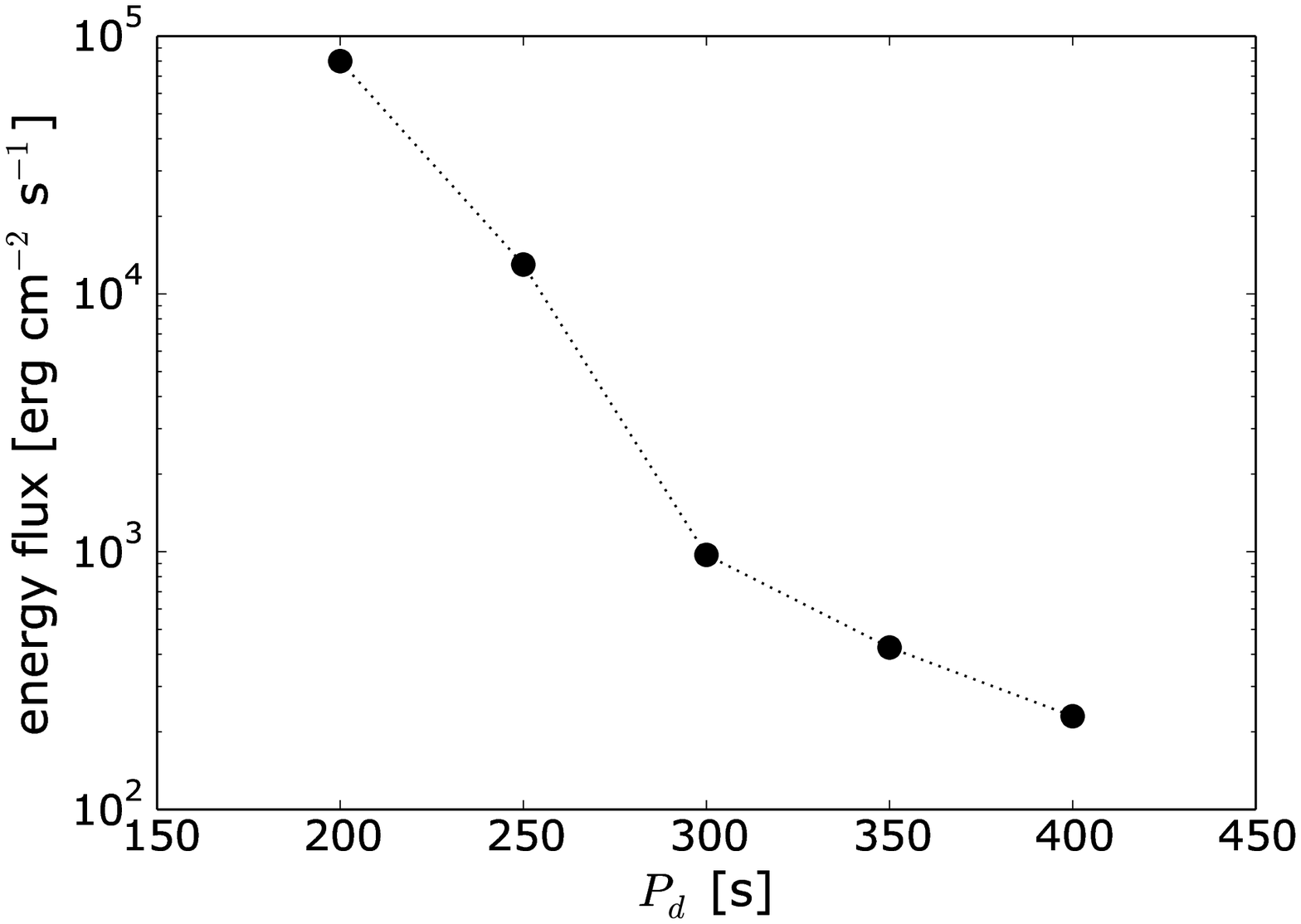}\\
\includegraphics[width=8cm]{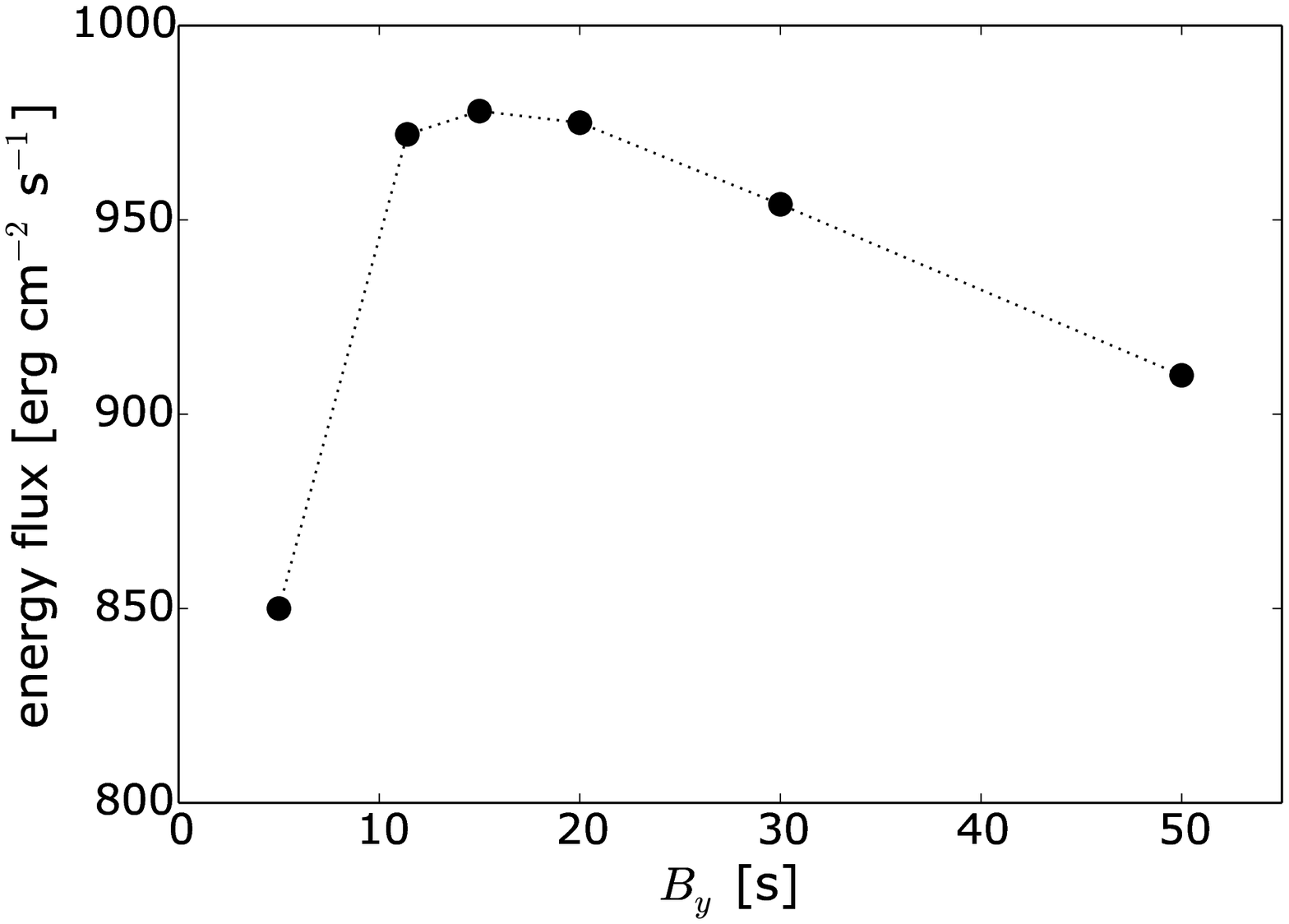}
\caption{\small  
Energy flux vs. the driving period for $B_{\rm y}=11.4$~G (top) 
and vs. magnetic field for $P_d=300$~s (bottom), evaluated at $y=1.9$~Mm.
}
\vspace{-0.5cm}
\label{fig:ef_Pd+By}
\end{center}
\end{figure}  
%
%
\subsection{Equilibrium}
We assume that the equilibrium plasma is still (${{\bf V}_i}={{\bf V}_n}={\bf 0}$) and the 
solar atmosphere is hydrostatic, viz. 
%
\begin{equation}
\varrho_{\rm {hi,n}}  g = -\frac{\partial p_{\rm {hi,n}}}{\partial y}\, .
\label{eq:33B}
\end{equation}
This equation is satisfied by
\begin{equation}
p_{\rm {hi,n}}(y) = p_{\rm {0i,n}} \exp\left[ -\int_{y_{\rm r}}^{y} \frac{dy'}{\Lambda_{\rm i,n}(y')} 
\right]\, , 
\end{equation}
\begin{equation}
\varrho_{\rm {hi,n}}(y) = \frac{p_{\rm {hi,n}}(y)}{g\Lambda_{\rm i,n}(y)}\,,
\end{equation}
where the subscript $_{\rm h}$ corresponds to a hydrostatic quantity, $ y_{\rm r}=50$~Mm is the reference level, $p_{\rm 0i}=10^{-2}$~Pa and $p_{\rm 0n}=3\cdot 10^{-4}$~Pa are, respectively, the ion and neutral gas pressure at this level  and 
\begin{equation}
\label{eq:lambda}
\Lambda_{\rm i,n}(y) =\frac{ k_{\rm B}T(y)}{ m_{\rm i,n}g}
\end{equation}
is the ion (neutral) pressure scale-height that depends on the plasma temperature $T(y)$,
which is taken from the semi-empirical model of Avrett \& Loeser (2008). See Fig.~\ref{fig:Atm} (top).  The ion and neutral mass density profiles at the equilibrium are shown in Fig.~\ref{fig:Atm} (bottom, solid line). Note that at $y=0$~Mm, which correspond to the bottom of 
the photosphere, $\varrho_n\approx 30\varrho_i$. For $y>0.7$~Mm $\varrho_i>\varrho_n$ and at $y=3$~Mm  $\varrho_i=80\varrho_n$.  The above hydrostatic equilibrium is overlaid by a current-free ($\nabla \times {\bf B}/\mu={\bf 0}$) magnetic field.
We consider the case of vertical magnetic field, ${\bf B}=[0, B_y, 0]$ with $B_y=11.4$~G. 
This value of magnetic field is typical for the solar corona 
and representative for the chromosphere. 

%
%
\subsection{Cutoff periods}
For the physical settings considered 
in this paper, the relevant cutoff periods were derived by Roberts (2004) for  acoustic waves, and by Roberts 
(2006) for MAWs. 
Comparison of the analytical findings derived in these papers 
to the numerical results presented here is given and discussed. 
The propagation of MAWs in the solar atmosphere is affected by cutoff periods for these waves. 
The local cutoff period derived for the vertical magnetic field by Roberts (2006) is defined as (see  Fig.~\ref{fig:Pm_cut_off})
\begin{equation}
P_m(y) = \frac{2\pi}{\Omega_m},
\label{eq:Pm}
\end{equation}
where
\begin{eqnarray}
\begin{split}
&\Omega_m^2(y)=\\
&c_t^2\left\{\frac{1}{4\Lambda^2}\left(\frac{c_t}{c_s}\right)^4-
\frac{1}{2}\gamma g \left(\frac{c_t^2}{c_s^4}\right)'+
\frac{1}{c_A^2}\left(\omega_g^2+\frac{g}{\Lambda}\frac{c_t^2}{c_s^2}\right)\right\}
\end{split}
\end{eqnarray}
and
$\omega_g^2$ denotes the squared buoyancy or Brunt-V\"{a}is\"{a}l\"{a}
 frequency given by 
\begin{equation}
\omega_g^2=-g\left(\frac{g}{c_s^2}+\frac{\varrho'}{\varrho}\right),
\end{equation}
with $c_t=c_s c_A / c_f$ being the 'cusp' (or tube) speed, $c_s = \sqrt{\gamma (p_{hi}+p_{hn}) / (\varrho_{hi}+\varrho_{hn})}$ sound speed, $c_A= B_0 /\sqrt{\mu (\varrho_{hi}+\varrho_{hn})}$ bulk Alfv\'en speed, 
$c_f(y)=\sqrt{c_s^2(y)+c_A^2(y)}$, 
$\Lambda=k_B T/(m_i+m_n)g$ and $'=d/dy$. 
\subsection{Periodic driver}
We assume that the waves are driven by the velocity fluctuations in the lower part of the photosphere.  Since 
typical periods of these fluctuations are about 5 min, we set a monochromatic driver at the height of $y=0$~Mm 
operating with the period close to $P=300$~s. We develop a numerical model for the wave propagation and this model generalizes the previous MHD studies of monochromatic MAWs waves performed by Kra\'skiewicz et al. (2019).

We perturb the solar atmosphere by the driver 
in the $y$-components of ion and neutral velocities given by
\begin{equation}
V_{\rm iy}(y=0,t)=
V_{\rm ny}(y=0,t)=
V_{\rm 0}\sin\left(\frac{2\pi\, t}{P_{\rm d}}\right),
\label{eq:2B}
\end{equation}
where $V_0=0.025$~km~s$^{-1}$ is the amplitude of the driver and $P_{\rm d}$ stands for its period. 
Here we consider wave-periods equal or longer than $200$~s. 
For such values of $P_{\rm d}$ collisions 
between ions and neutrals are not expected to play 
any significant role in wave damping and consequently in thermalization 
of their energy. 
However, frequent ion-neutral collisions in 
the lower atmospheric layers result in strong 
ion-neutral 
coupling in the photosphere and less frequent collisions in weak coupling 
in the top chromosphere and the low corona, 
which affects cutoff wave-periods. 

The driver generates neutral acoustic and ion MAWs waves propagating through the photosphere and then through the chromosphere towards the corona. We study 
these waves by tracing the $y$-components of ion and neutral velocities and their mass densities.  We aim is to show that some waves of their wave-periods larger than 300~s are evanescent 
and they are not able to reach the solar corona.  Note that internal gravity waves are removed from the model by considering the background medium such as 1D plasma, which means that $\partial/\partial x= \partial/\partial z=0$.
%
\section{NUMERICAL RESULTS }\label{sec:num_sim_model}
%
We perform numerical simulations of the MAWs in a partially ionized and weakly 
magnetized solar atmosphere by using the JOANNA code (W\'ojcik et al. 
2018).  
The code is based on the original work by Leake et al. (2012), who 
considered a two-fluid model of ions and neutrals as separate fluids.  
This code was extensively tested in W\'ojcik et al. (2019, 2020) and in Murawski, Musielak \& W\'ojcik (2020); the latter paper reports on a good agreement between the results 
obtained by the code and the observations.  

We present results of numerical simulations with use of a 1D
uniform grid of its finest size of $\Delta y=5$~km between $y = 0$ and 
$y = 5.12$~Mm.  Higher up the grid stretches up with height until the top 
boundary which is located at $y = 60$~Mm.  At the bottom and top boundaries, 
we set and hold fixed in time all plasma quantities to their equilibrium values.
The only exception is the bottom boundary at which we overlay the equilibrium 
quantities by the driver, specified by Eq. (\ref{eq:2B}).  This means that the 
presented results are valid for the photosphere and chromosphere, and that 
they can be used to establish ranges of waves that may carry energy up to 
the corona. 

We focus on waves with periods of a few hundred seconds because for these 
waves the ion-neutral collisions are ineffective in thermalization of the wave 
energy.  On the other hand, if wave periods are an order of magnitude shorter,
then their periods become close to the ion-neutral collision time (e.g., Popescu 
Braileanu et al. 2019), and therefore these waves are not considered here.  
Since waves of short periods are mainly responsible for the heating of the 
solar chromosphere, the latter is not discussed in this paper, but instead the 
presented results are focused on the wave energy transport to the solar corona.
In the following, we describe in detail the obtained numerical results and discuss
their implications for solar physics.  
\subsection{The case of $P_{\rm d}=200$~s}
In Fig.~\ref{fig:200_Viy} (top), we show the time-distance plot for the ion vertical velocity, 
$V_{iy}$, for the wave-period, $P_{\rm d}=200$~s.  It is seen that the signal evolution 
remains essentially unchanged in time for a fixed height, and that the wave signals propagate 
through the photosphere and chromosphere to the corona.   Figure~\ref{fig:200_Viy} (bottom) 
demonstrates the Fourier period, $P$ vs. height, $y$ in which we clearly see the main period 
$P_d=200$~s and also $P_{\rm cutoff}=230$~s, albeit with less power, for $1.0\,{\rm Mm}<y<1.5\,
{\rm Mm}$. The same period we can infer from the formula $P_{\rm cutoff}=
P_{\rm mod}P_{\rm d}/(P_{\rm mod}-P_{\rm d})$, where the  modulation period, $P_{\rm mod}\approx 1470$~s, 
is estimated from Fig.~\ref{fig:200_wavelet_Viy-1.5} ((a) panel). The period of 230~s is consistent with the cutoff 
wave-period for MAWs at $y=1.5$~Mm  (Fig.~\ref{fig:Pm_cut_off}), and it is exited 
together with the driver period.

In Fig.~\ref{fig:200_wavelet_Viy-1.5}, we demonstrate the time-signature of $V_{iy}(y=1.5$~Mm) 
(top) and the corresponding wavelet spectra (bottom) for the wave-period $P_{\rm d}=200$~s.  The time-signature exhibits the modulations with $P_{\rm mod}$ (top), which is discernible in the wavelet spectrum for 
$70$~s < $t$ < $590$~s (bottom-left).
The presented results reinforce those shown in Fig.~\ref{fig:200_Viy}.  The main implication of 
these results is that the propagating wave pattern is seen for the waves of this period in the solar 
photosphere and chromosphere, which means that these waves carry most of their energy directly 
to the solar corona. 

Let us point out that as in the case of $P_d=200$~s the differences between the neutral and ion velocity periodograms 
are very small, we show here only the ion case.
%
%
%
%
\subsection{The case of $P_{\rm d}=300$~s}
Figure~\ref{fig:300_Viy} presents time-distance plot for $V_{iy}$ (top) and its Fourier 
spectrum (bottom) for $P_d=300$~s.  The presented results show that the waves with 
periods of $P=300$~s cannot reach the solar corona as their power spectrum significantly 
decreases with height in the middle chromosphere.  Moreover, it is also seen that shorter period 
waves appear. These waves correspond to the local cutoff with the strongest signal associated with $P_{\rm cutoff}\approx 240$~s, observed at $y>1$~Mm 
in Figs.~\ref{fig:Pm_cut_off} and \ref{fig:300_Viy}. 
A comparison of these wave periods with the observational data of Wi\'sniewska et al. (2016) and Kayshap et al. (2018) reveals an agreement at some points. 
Hence, the qualitative similarity between the numerical results and 
the observational findings confirms that ion-neutral collisions are 
an efficient mechanism of energy release. 

Figure~\ref{fig:300_wavelet_Viy-1.5} illustrates time-signature for $V_{iy}(y=1.5$~Mm) 
(top) and the corresponding wavelet spectra (bottom).  Despite different wave periods 
displayed in Figs.~\ref{fig:200_wavelet_Viy-1.5} and \ref{fig:300_wavelet_Viy-1.5}, the 
wave evolution remains apparently unchanged. However, the wave profile is altered in comparison 
to the driving signal, with clear signs of modulation seen in the time-signature (top).  The 
modulation period, $P_{\rm mod}$, in this case can be estimated from Fig.~\ref{fig:300_wavelet_Viy-1.5} 
(top) as $P_{\rm mod}\approx 1200$~s, giving $P_{\rm cutoff}=P_{\rm mod}P_{\rm d}/
(P_{\rm mod}+P_{\rm d})\simeq 240$~s. This value is close to 230~s which was estimated 
for $P_d=200$~s and it is present for $y>1$~Mm in Fig.~\ref{fig:Pm_cut_off}.
\subsection{The case of $P_{\rm d}=400$~s}
Figure~\ref{fig:400_Viy} presents time-distance plot for $V_{iy}$ (top)  and its Fourier 
power spectrum (bottom) for $P=400$~s. Comparison to the results of previous figures 
demonstrates that these waves can reach only the atmospheric height $y\simeq 1.3$ Mm. 
As a result, these waves cannot become propagating until they reach the solar corona, and 
the waves with period of $P_{\rm cutoff}\approx 255$~s are excited at about $y=1.2$~Mm 
(bottom). This period is close to cutoff periods seen in Fig.~\ref{fig:Pm_cut_off} for $y>1$~Mm.

Note that the results presented in Fig.~\ref{fig:400_Viy} (bottom) reveal at the top 
chromosphere the wave-periods which are much lower than the driving period of $P_{\rm d}
=400$~s.  These wave-periods are in the range of $200-300$~s and they support the results 
given by Fleck \& Schmitz (1991) and Kalkofen et al. (1994).  According to these authors, 
any $P_{\rm d}$ higher than the local atmospheric cutoff period initially oscillates with its 
local cutoff and later on the oscillations evolve to reach $P_{\rm d}$ (see Fig.~\ref{fig:wavelet_Pd400-1.5},  left-bottom). The global wavelet spectrum reveals that the signal 
at $P=P_d=400$~s is much lower than the signal at $P\approx 230$~s (right-bottom), 
which agrees with Fig.~\ref{fig:400_Viy} (bottom) which shows that for $y>1.5$~Mm 
the signal corresponding to $P\simeq 230$~s is stronger than that for $P=400$~s.

As expected from our previous results obtained for the waves with periods of 200 s
and 300 s, the waves with their periods of 400 s cannot reach the solar chromosphere
because their power spectrum is significantly reduced in the solar chromosphere and 
its shape is altered, so the waves become evanescent.  By using the latter wave property,
we were able to determine the cutoff period for the waves in the solar chromosphere 
and corona, and use it to establish the conditions for the wave propagating in these 
layers of the solar atmosphere. 
\subsection{Parametric studies} 
Figure~\ref{fig:cutoff_Pd} displays main Fourier period for 
$V_{\rm iy}$ vs. maximum value of $y$ at which this period is present, 
i.e. $P_{\rm m}=P_{\rm d}$ for $y < max(y)$; 
for $y > max(y)$ the dominant period in the Fourier spectrum is lower 
than the driving period $P_{\rm d}$. 
Note the fast fall-off of $P_{\rm m}$ in the range of 
$0.75\, {\rm Mm} < y < 1\, {\rm Mm}$ which means than 
the driver with its wave-period $P_{\rm d}$ higher than about 
$300\,{\rm s}$ excites evanescent magnetoacoustic waves 
reaching the low chromosphere. 
Such a drastic decline of wave wave-periods was already detected 
in the numerical data for stochastically excited magnetoacoustic-gravity waves in a quiet region of the solar atmosphere, see Fig.~4 in  Kra\'skiewicz and Murawski (2019), 
%
and in the observational data of Wi\'sniewska et al. (2016). 

We discuss now the energy flux which is approximated 
by the following formula: 
\begin{equation}
F_{\rm E} \approx \frac{1}{2} \varrho_{\rm i} \mathbf{V}_{\rm i}^2 \, c_{\rm s}\, .
\label{eq:energy_flux}\\
\end{equation}
%
%
%
We evaluate this flux at $y=1.9$~Mm. 
Figure~\ref{fig:ef_Pd+By} illustrates $F_{\rm E}$ vs. 
the driving period, $P_{\rm d}$, (top) and magnetic field, 
$B_{\rm y}$, (bottom). 
Note that $F_{\rm E}$ declines with $P_{\rm d}$, meaning that 
shorter wave-period waves carry more energy than longer wave-period waves (top). 
The variation of $F_{\rm E}$ with $B_{\rm y}$ reveals that for 
$B_{\rm y}=5$~G 
$F_{\rm E}\approx 850\, {\rm erg\, cm}^{-2}\, {\rm s}^{-1}$. 
For higher values of $B_{\rm y}$ $F_{\rm E}$ grows, attaining its 
local maximum at $B_{\rm y}=15$~G, and then 
$F_{\rm E}$ declines to $F_{\rm E}\approx 910$~G for $B_{\rm y}=50$~G 
(bottom). 
These energy fluxes can be compared with the data in 
Tab.~1 from Withbroe \& Noyes (1977)
according to which total energy loss is 
$3\cdot 10^5\, {\rm erg\, cm}^{-2}\,{\rm s}^{-1}$. 
The data of Fig.~\ref{fig:ef_Pd+By} shows that 
$F_{\rm E}$ is close to 
$1\cdot 10^5\, {\rm erg\, cm}^{-2}\,{\rm s}^{-1}$ 
for $P_{\rm d}=200$~s and $B_{\rm y}=11.4$~G (top). 
As for other considered values $F_{\rm E}$ is lower than 
the required values to balance energy losses in the top chromosphere 
(Withbroe \& Noyes 1977), we conclude that only the monochromatic 
driver with its period $P_{\rm d}=200$~s and $B_{\rm y}=11.4$~G 
is able to provide the required energy flux to heat the top chromosphere. 
%
%
%
\subsection{Oscillations in mass densities}
It is noteworthy that in the limit of long wave-period 
waves, that is discussed in this paper, ions and neutrals acquire similar velocities, $\mathbf{V}_{\rm i}\simeq
\mathbf{V}_{\rm n}$. 
As a result of that the ion-neutral drag force is negligible and the 
collisional heating is marginal.  Besides, wave dispersion is minimal (e.g., Zaqarashvili et al. 
2011) and the two-fluid equations approach is limited to the two-species equations (e.g. 
Terada et al. 2009). These two-species equations reveal potentially different evolution of ion and neutral mass densities. Therefore, we discuss this evolution here.

\begin{figure} 
\begin{center}
\includegraphics[width=8cm]{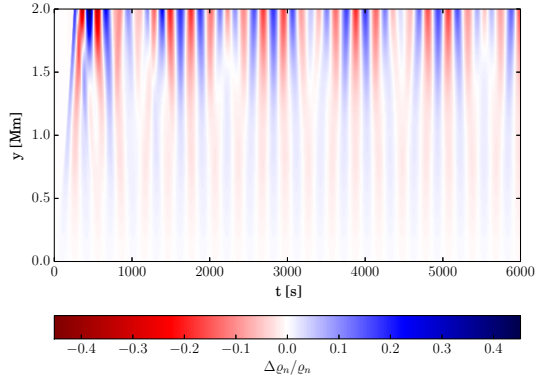}
\caption{\small Time-distance plots for $\Delta\varrho_{\rm n}/\varrho_{\rm {n}}$ and $P_{\rm d}=300$~s.
}
\label{fig:delta_Rho_i_n}
\end{center}
\end{figure}

\begin{figure} 
\begin{center}
\includegraphics[width=8cm]{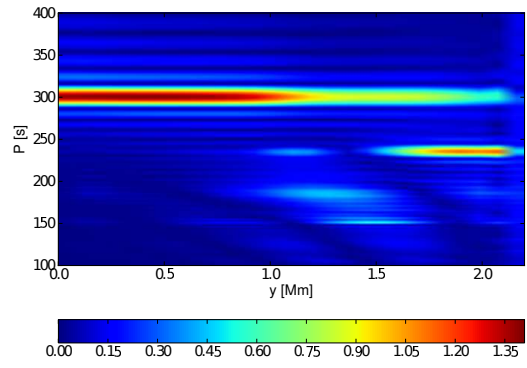}
\includegraphics[width=8cm]{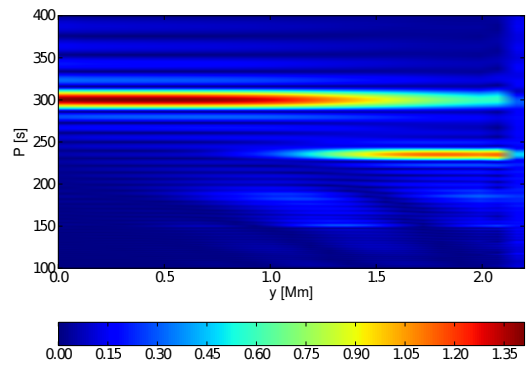}
\caption{\small Fourier period, $P$, vs. height, $y$, for $\varrho_{i}$ (top) and $\varrho_{n}$ (bottom) in the case of $P_{\rm d}=300$~s.}
\vspace{-0.5cm}
\label{fig:Map_300_Rho}
\end{center}
\end{figure}

\begin{figure*} 
\begin{center}
\includegraphics[width=14cm, height=11cm]{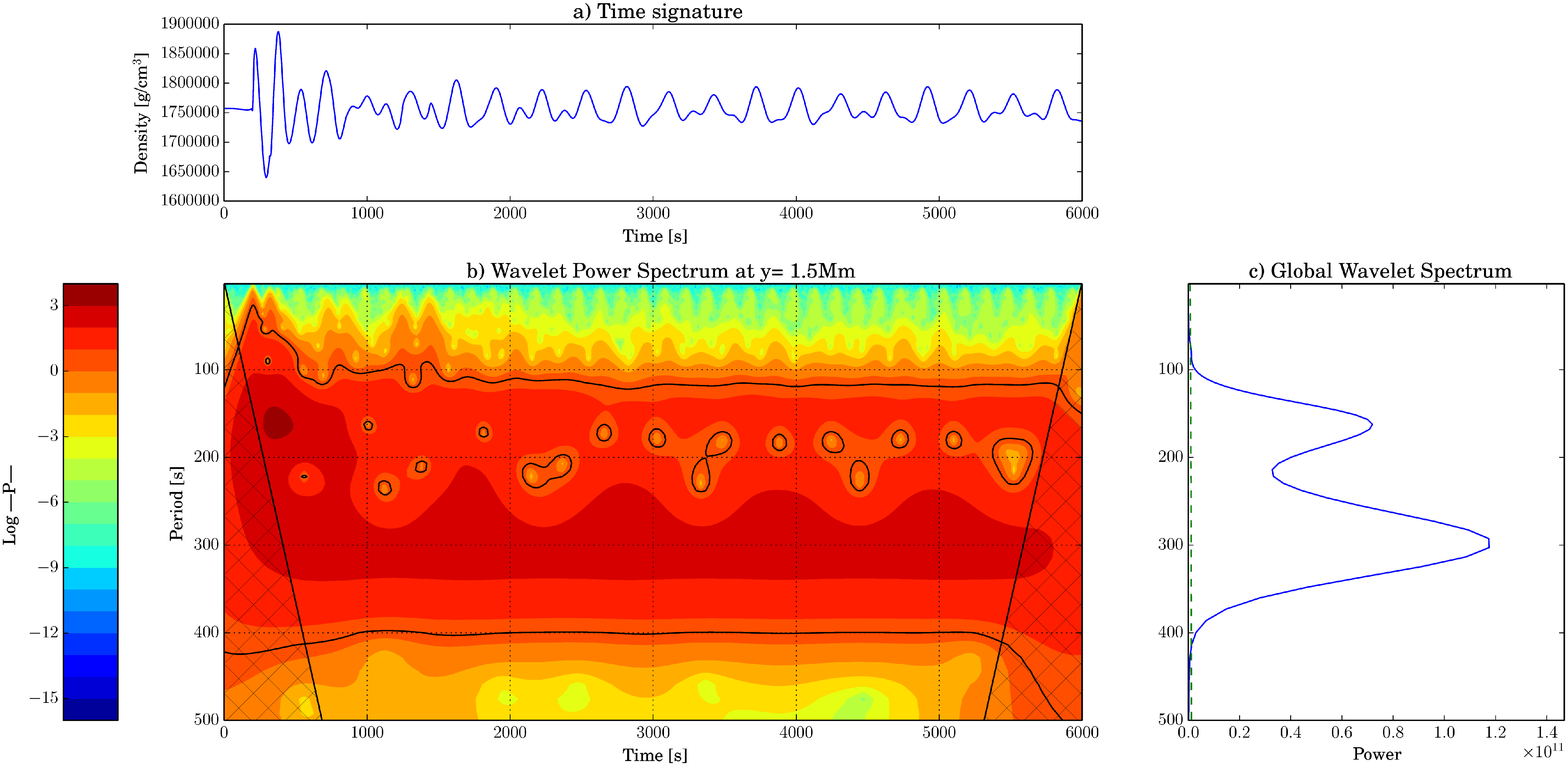}
\includegraphics[width=14cm, height=11cm]{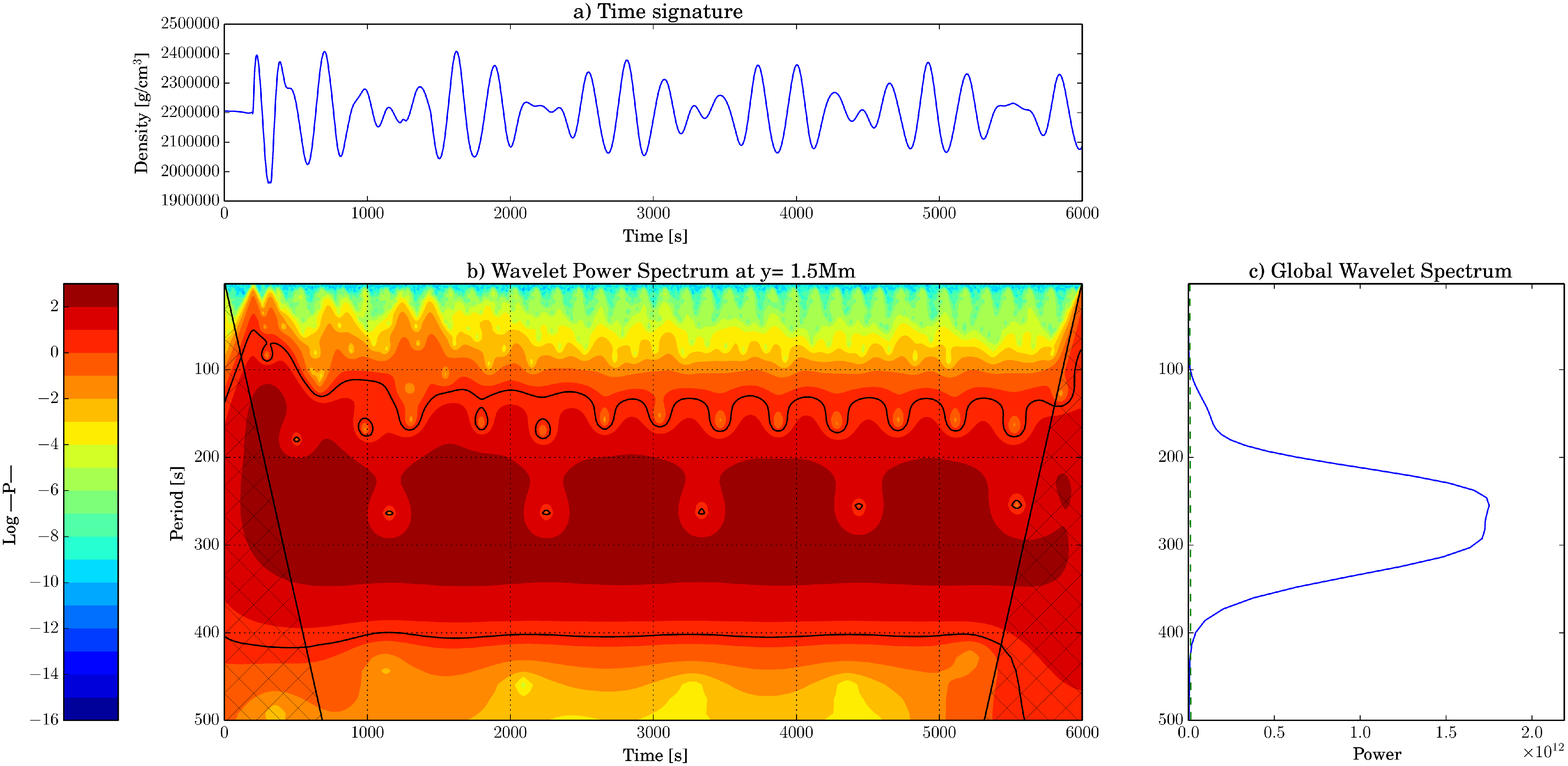}
\caption{\small Wavelet spectrum for $\varrho_{i}$ (top) and $\varrho_{n}$ (bottom), collected 
at $y=1.5$~Mm, in the case of $P_{\rm d}=300$~s.}
\label{fig:wavelet_300_Rho}
\end{center}
\end{figure*}

Figure~\ref{fig:delta_Rho_i_n} describes the time-distance plot for relative perturbations of 
the neutral mass density, $\Delta\varrho_{\rm n}/\varrho_{\rm {n}}=(\varrho_{\rm {n}}-
\varrho_{\rm {hn}})/\varrho_{\rm {n}}$, from the equilibrium state. The pattern is similar 
to $V_{iy}(t,y)$  but the investigation of the Fourier power spectrum shows some different 
behaviour of ions from neutrals ({Fig.~\ref{fig:Map_300_Rho}}).  More significant difference 
is seen in Fig.~\ref{fig:wavelet_300_Rho} which shows time-signatures for $\varrho_i$ and 
$\varrho_n$, collected at $y=1.5$~Mm (a panels), the corresponding wavelet spectra 
(left-bottom panels), and the global wavelet spectra (right-bottom panels). These global 
wavelet spectra reveal the excited signal of $P=150$~s for ions (top) and lack of that 
for neutrals (bottom panel) at $y=1.5$~Mm. These different oscillations of $\varrho_{\rm {i}}$ 
and $\varrho_{\rm {n}}$ from those in $V_{iy}$ are not surprising as even in the case of linear MHD 
mass density is governed by a different evolution equation than vertical velocity is (Roberts 
2006).

The presented results show that the waves with periods of 300 s cannot reach the solar 
corona because of their spectrum decreases significantly in the middle chromosphere.
This clearly shows the effects of the solar chromosphere on the waves of this period,
which is very different than the effects of the chromosphere on the waves with periods
of 200 s.  Our results also demonstrate the changes between the driven and propagating 
wave profiles in the solar chromosphere.  We also report on differences in the behaviour 
of ions and neutrals, which are revealed in their different oscillations, but we point out 
that these differences have already been observed before. 
%


%
%

%
%
\section{CONCLUSIONS AND SUMMARY}\label{sec:conclusions}
%
We performed 1D numerical simulations of propagation of neutral acoustic and ion 
magnetoacoustic two-fluid waves in the partially ionized lower solar atmosphere, 
consisting of ion+electron and neutral fluids, which are coupled by ion-neutral 
collisions.  
The considered atmosphere was assumed to be 
permitted by a uniform vertical magnetic field 
and the waves were excited by a 
monochromatic driver located at the bottom of the solar photosphere.  
We 
investigated variations of wave-periods with height in the solar atmosphere, 
which allowed us to find the wave cutoffs and determine the conditions for the 
wave propagation.  

Our main results demonstrate that waves with different wave-periods than those
long wave-periods generated by the driver are present and that their vertical propagation is strongly
affected by the solar atmosphere, which filters out the $300$~s (and longer) waves and makes some of these waves evanescent.  
By identification the evanescent waves we were able to
find the wave cutoffs and their variations in the solar atmosphere, 
and use them to determine the propagation conditions for 
the considered waves. 
Then, we used the propagation conditions to find periods of waves 
that may carry their energy from the solar surface to the corona.
Additionally, we showed that according to the analytical predictions 
(e.g. Roberts 2006) oscillations in mass densities exhibit different wave period spectra than in their vertical velocity component. 

Finally we mention that a two-fluid model describes dynamics of two fluids, such as ions+electrons and neutrals. 
An MHD model does not distinguish between these two fluids and 
therefore it is inferior to a two-fluid model. Our results reveal that Fourier 
spectra of excited oscillations of these two fluids differ one from another and 
they depend on whether these spectra 
are drawn for vertical velocities (see Sect. 3) or for mass densities 
(see Appendix). 
Obviously for wave-periods much larger 
than ion-neutral collision time the spectra for vertical velocities 
are similar in the framework of MHD and 
the two-fluid model. 

\bigskip\noindent
{\bf ACKNOWLEDGMENTS}\\
\noindent 
The JOANNA code was developed by Darek W\'{o}jcik 
with 
some contribution from Luis Kadowaki and Piotr Wo{\l}oszkiewicz. 
KM's work was done within the framework of the projects from the Polish Science Center
(NCN) Grant No. 
2020/37/B/ST9/00184. 
%


\bigskip\noindent
{\bf DATA AVAILABILITY}\\
The data underlying this article are available in the article and in its online supplementary material.

\bigskip\noindent
{\bf REFERENCES}\\
\noindent 
Avrett, E. H. \& Loeser, R. 2008, ApJS, 175, 229\\
Ballester, J.~L., Alexeev, I., Collados, M., {et~al.}\ 2018, Space Science
Reviews, 214, 58B\\
Cally, P.~S., \& Hansen, S.~C.\ 2011, ApJ, 738, 119\\
Chmielewski, P., Srivastava, A.~K., Murawski, K., et al.\ 2013, MNRAS, 428, 40\\
Defouw, R.~J.\ 1976, ApJ, 209, 266\\
Felipe, T., Kuckein, C., \& Thaler, I.\ 2018, A\&A, 617, A39\\
Fleck, B., \& Schmitz, F.\ 1991, A\&A, 250, 235\\
Fleck, B., \& Schmitz, F.\ 1993, A\&A, 273, 671\\
Gough D. O.\ 1977, ApJ, 214, 196\\
Kalkofen W., Rossi P., Bodo G., Massaglia S.\ 
1994, A\&A, 284, 976\\
Kayshap, P., Murawski, K., Srivastava, A.~K., et al.\ 2018, MNRAS, 479, 5512\\
Kra\'skiewicz, J.K. \& Murawski, K.\ 2019, MNRAS, 482, 3244\\
Kra\'skiewicz, J.K., Murawski, K., \& Musielak, Z.E.\ 2019, A\&A, 623, A62\\
{Ku{\'z}ma}, B., {W{\'o}jcik}, D., {Murawski}, K.\ 2019, A\&A, 878, 81\\ 
{Ku{\'z}ma}, B., {Murawski}, K., \& Musielak, Z.E.\ 2022, submitted to MNRAS\\
Lamb, H.\ 1909, Proc. Lond. Math. Soc., 7, 122\\
Lamb, H.\ 1910, Proc. R. Soc. London, A, 34, 551\\ 
Lamb, H.\ 1945, Hydrodynamics, Dover Publications, New York\\
Leake, J. E., Lukin, V. S., Linton, M. G., Meier, E. T.\ 2012, ApJ, 760, 109\\
Mart{\'{\i}}nez-Sykora, J., Pontieu, B.~D., Hansteen, V.~H., {et~al.} 2017,
Science, 356, 1269\\ 
Murawski, K., Musielak, Z.~E., Konkol, P., et al.\ 2016, ApJ, 827, 37\\
Musielak, Z.~E., Musielak, D.~E., \& Mobashi, H. \ 2006, Phys. Rev. E, 73, 036612-1\\
Maneva, Y.G., Alvarez Laguna, A., Lani, A., \& Poedts, S.\ 2017, ApJ, 836, 197\\
Murawski, K., \& Zaqarashvili, T. V.\ 2010, A\&A, 519, 8\\ 
Murawski, K., \& Musielak, Z.~E.\ 2010, A\&A, 518, A37\\
Murawski, K., \& Musielak, Z.E.\ 2016, MNRAS, 463, 4433\\ 
Murawski, K., Musielak, Z.E., W\'ojcik, D.\ 2020, ApJL, 896 L1\\ 
Musielak, Z. E.\ 1990, ApJ, 351, 287\\
Musielak Z. E., Moore R., 1995, ApJ , 452, 434\\
Musielak, Z.~E., Routh, S., \& Hammer, R.\ 2007, ApJ, 659, 650\\
Musielak Z. E., Musielak, D. E., Mobashi H.\ 2006, Phys. Rev. E, 73, 036612-1\\
Narain, U., \& Ulmschneider, P.\ 1996, Space Sci. Rev., 75, 453\\
Oliver, R., Soler, R., Terradas, J., \& Zaqarashvili, T. V. 2016, ApJ,818, 128\\
Perera, H.~K., Musielak, Z.~E., \& Murawski, K.\ 2015, MNRAS, 450, 3169\\
Priest, E. R.\ 2014, Magnetohydrodynamics of the Sun, Cambridge University Press, Cambridge\\
Popescu Braileanu, B., Lukin, V.~S., Khomenko, E., et al.\ 2019, A\&A, 627, A25\\
Rae, I. C., \& Roberts, B.\ 1982, ApJ, 256, 761\\
Roberts, B.\ 1991, Geophysical and Astrophysical Fluid Dynamics, 62, 83\\
Roberts, B., \& Ulmschneider, P.\ 1997, European Meeting on Solar Physics, 75\\
Roberts, B.\ 2004, Proceedings of 'SOHO 13 Waves, Oscillations and Small-scale 
Transients Events in the Solar Atmosphere: Joint View from SOHO and TRACE', 
Compiled by: H. Lacoste, p. 1\\
Roberts, B.\ 2006, Phil. Trans. R. Soc. A, 364, 447\\
Rogers, F. J. \& Nayfonov, A. \ 2002, ApJ, 576, 1064\\
Routh S., Musielak Z. E. 2014, Astronomische Nachrichten, 335, 1043\\
Routh, S., Musielak, Z.~E., \& Hammer, R.\ 2010, ApJ, 709, 1297\\
Routh, S., Musielak, Z.E., Sundar, M.N., Joshi, S.S., \& Charan, S.\ 2020, Astrohys. \& 
Space Sci., 365, 139\\
Schmitz, F., \& Fleck, B.\ 1992, A\&A, 260, 447\\
Stark, B.A., \& Musielak, Z.E.\ 1993, ApJ, 409, 450\\
Terada, N., Shinagawa, H., Tanaka, T., Murawski, K.,  Terada, K.\ 2009, JGR, 114, A09208\\
V{\"o}gler, A. 2004, Three-dimensional simulations of magnetoconvection
in the solar photosphere (Copernicus)\\
Vranjes, J., \& Krstic, P. \ 2013, App, 554, A22\\
Wi{\'s}niewska, A., Musielak, Z.~E., Staiger, J., Roth M.\ 2016, ApJ, 819, L23\\
Withbroe, G. L. \& Noyes, R. W.\ 1977, Annual Rev. Astr. Astrophys., 15. (A78-16576 04-90) Palo Alto, Calif., Annual Reviews, Inc., 1977, p. 363\\
{W{\'o}jcik}, D., {Murawski}, K., \& {Musielak}, Z.~E. 2018, MNRAS, 481, 262\\ 
W\'ojcik, D., Murawski, K., \& Musielak, Z.E.\ 2019, ApJ, 882, 32W\\ 
W\'ojcik, D., Ku\'zma, B., Murawski, K., Musielak, Z.E.\ 2020, A\&A, 635, A28\\ 
Zaqarashvili, T.V., Khodachenko, M.L., Rucker, H.O.\ 
2011, A\&A, 529, A82\\

\end{document}